\documentclass[fleqn,usenatbib,useAMS]{mnras}

\usepackage{graphicx} 
\graphicspath{ {./images/} }
\usepackage{amsmath} 
\usepackage{multicol}        
\usepackage{bm} 
\usepackage{booktabs}
\usepackage{anyfontsize}

\usepackage[T1]{fontenc}
\usepackage{ae,aecompl}

\usepackage{newtxtext,newtxmath}
\usepackage{orcidlink}

\title[TOI-1117]{The TOI-1117 Multi-planetary System: 3 sub-Neptunes, 1 in both the Neptunian Desert and Radius Valley}

\author[I. S. Lockley]{Isobel S. Lockley$^{\orcidlink{0009-0003-0928-3588}}$,$^{1,2}$\thanks{E-mail: \href{mailto:isobel.lockley@warwick.ac.uk}{isobel.lockley@warwick.ac.uk}}
David J. Armstrong$^{\orcidlink{0000-0002-5080-4117}}$,$^{1,2}$
Jorge Fern\'{a}ndez Fern\'{a}ndez,$^{1,2}$
Sarah Millholland$^{\orcidlink{0000-0003-3130-2282}}$,$^{3,4}$
\newauthor
Henrik Knierim$^{\orcidlink{0009-0003-0637-9170}}$,$^{5}$
Elisa Delgado Mena,$^{6,7}$
Sergio Sousa$^{\orcidlink{0000-0001-9047-2965}}$,$^{7}$
Karen A.\ Collins$^{\orcidlink{0000-0001-6588-9574}}$,$^{8}$
Cristilyn N.\ Watkins$^{\orcidlink{0000-0001-8621-6731}}$,$^{8}$
\newauthor
Steve~B.~Howell$^{\orcidlink{0000-0002-2532-2853}}$,$^{9}$
Vardan Adibekyan$^{7}$
Ravit Helled$^{\orcidlink{0000-0001-5555-2652}}$,$^{5}$
Carl Ziegler$^{\orcidlink{0000-0002-0619-7639}}$,$^{10}$
Daniel Bayliss$^{\orcidlink{0000-0001-6023-1335}}$,$^{1,2}$
\newauthor
C\'{e}sar Brice\~{n}o,$^{11}$
Amadeo Castro-Gonz\'{a}lez$^{\orcidlink{0000-0001-7439-3618}}$,$^{12}$
Catherine~A.~Clark$^{\orcidlink{0000-0002-2361-5812}}$,$^{13}$
Kevin I.\ Collins$^{\orcidlink{0000-0003-2781-3207}}$,$^{14}$
\newauthor
Jessie L. Christiansen$^{\orcidlink{0000-0002-8035-4778}}$,$^{13}$
Kaiming Cui$^{\orcidlink{0000-0003-1535-5587}}$,$^{1,2}$
Rodrigo Diaz$^{\orcidlink{0000-0001-9289-5160}}$,$^{15}$
Jon M. Jenkins$^{\orcidlink{0000-0002-4715-9460}}$,$^{9}$
Marcelo A. F. Keniger$^{\orcidlink{0009-0005-2761-9190}}$,$^{1,2}$
\newauthor
Michelle Kunimoto$^{\orcidlink{0000-0001-9269-8060}}$,$^{16}$
Nicholas Law,$^{17}$
Jorge Lillo-Box$^{\orcidlink{0000-0003-3742-1987}}$,$^{12}$
Colin Littlefield$^{\orcidlink{0000-0001-7746-5795}}$,$^{9,18}$
Andrew W. Mann$^{\orcidlink{0000-0003-3654-1602}}$,$^{17}$
\newauthor
Morgan A. Mitchell$^{\orcidlink{0009-0004-6130-7775}}$,$^{1,2}$
Louise D. Nielsen$^{\orcidlink{0000-0002-5254-2499}}$,$^{19}$
Jos\'e Rodrigues$^{\orcidlink{0000-0001-5164-360}}$,$^{7}$
Pam Rowden,$^{20}$
Nuno C. Santos$^{\orcidlink{0000-0003-4422-2919}}$,$^{7,21}$
\newauthor
Sara Seager$^{\orcidlink{0000-0002-6892-6948}}$,$^{3,4}$
Peter J. Wheatley$^{\orcidlink{0000-0003-1452-2240}}$,$^{1,2}$ and 
Joshua Winn$^{\orcidlink{0000-0002-4265-047X}}$ $^{22}$
\\\\
$^{1}$Department of Physics, University of Warwick, Gibbet HIll Road, Coventry CV4 7AL, UK\\
$^{2}$Centre for Exoplanets and Habitability, University of Warwick, Gibbet Hill Road, Coventry CV4 7AL, UK\\
$^{3}$Department of Physics, Massachusetts Institute of Technology, Cambridge, MA 02139, USA\\
$^{4}$MIT Kavli Institute for Astrophysics and Space Research, Massachusetts Institute of Technology, Cambridge, MA 02139, USA\\
$^{5}$Department of Astrophysics, University of Zurich, Winterthurerstr. 190, CH-8057 Zurich, Switzerland\\
$^{6}$Centro de Astrobiolog\'ia (CAB), CSIC-INTA, ESAC campus, Camino Bajo del Castillo s/n, 28692, Villanueva de la Ca\~nada (Madrid), Spain\\
$^{7}$Instituto de Astrof\'isica e Ci\^encias do Espa\c{c}o, Universidade do Porto, CAUP, Rua das Estrelas, 4150-762 Porto, Portugal\\
$^{8}$Center for Astrophysics \textbar \ Harvard \& Smithsonian, 60 Garden Street, Cambridge, MA 02138, USA\\
$^{9}$NASA Ames Research Center, Moffett Field, CA 94035, USA\\
$^{10}$Department of Physics, Engineering and Astronomy, Stephen F. Austin State University, 1936 North St, Nacogdoches, TX 75962, USA\\
$^{11}$Cerro Tololo Inter-American Observatory, Casilla 603, La Serena, Chile\\
$^{12}$Centro de Astrobiolog\'ia (CAB), CSIC-INTA, ESAC campus, Camino Bajo del Castillo s/n, 28692, Villanueva de la Ca\~nada (Madrid), Spain\\
$^{13}$NASA Exoplanet Science Institute, IPAC, California Institute of Technology, Pasadena, CA 91125 USA\\
$^{14}$George Mason University, 4400 University Drive, Fairfax, VA, 22030 USA\\
$^{15}$International Center for Advanced Studies (ICAS) and ICIFI (CONICET), ECyT-UNSAM, Campus Miguelete, 25 de Mayo y Francia, (1650) Buenos Aires, Argentina.\\
$^{16}$Department of Physics and Astronomy, University of British Columbia, 6224 Agricultural Road, Vancouver, BC V6T 1Z1, Canada\\
$^{17}$Department of Physics and Astronomy, The University of North Carolina at Chapel Hill, Chapel Hill, NC 27599-3255, USA\\
$^{18}$Bay Area Environmental Research Institute, Moffett Field, CA 94035, USA\\
$^{19}$University Observatory, Faculty of Physics, Ludwig-Maximilians-Universit{\"a}t M{\"u}nchen, Scheinerstr. 1, 81679 Munich, Germany\\
$^{20}$Royal Astronomical Society, Burlington House, Piccadilly, London W1J 0BQ, UK\\
$^{21}$Departamento de F\'isica e Astronomia, Faculdade de Ci\^encias, Universidade do Porto, Rua do Campo Alegre, 4169-007 Porto, Portugal\\
$^{22}$Department of Astrophysical Sciences, Princeton University, Princeton, NJ 08544, USA
}

\date{XXX}

\pubyear{2024}

\begin{document}
\label{firstpage}
\pagerange{\pageref{firstpage}--\pageref{lastpage}}
\maketitle

\begin{abstract}
We present the discovery of three sub-Neptune planets around TOI-1117, a Sun-like star with mass $0.97\pm0.02M_{\odot}$, radius $1.05\pm0.03R_{\odot}$, age $4.42\pm1.50$ Gyr and effective temperature $5635\pm62$ K. Light curves from TESS and LCOGT show a transiting sub-Neptune with a $2.23$ day period, mass $M_b=8.90_{-0.96}^{+0.95}M_{\oplus}$ and radius $R_b=2.46_{-0.12}^{+0.13}R_{\oplus}$. This is a rare 'hot Neptune' that falls within the parameter spaces known as the ‘Neptunian Desert' and the ‘Radius Valley'. Two more planetary signals are detected in HARPS radial velocities, revealing two non-transiting planets with minimum masses $M_c=7.46_{-1.62}^{+1.43}M_{\oplus}$ and $M_d=9.06_{-1.78}^{+2.07}M_{\oplus}$, and periods of $4.579\pm0.004$ and $8.67\pm0.01$ days. The eccentricities were poorly constrained by the HARPS data, with upper limits $e_b=0.11$, $e_c=0.29$, and $e_d=0.24$. However, dynamical simulations of the TOI-1117 system, suggest that the orbits must be nearly circular to be stable. The simulations also show that TOI-1117\,b and c are likely to be in a near 2:1 resonance. The multi-planet nature of TOI-1117 makes it a more complex case for formation theories of the Neptunian Desert and Radius Valley, as current theories such as high-eccentricity migration are too turbulent to produce a stable, non-eccentric, multi-planet system. Moreover, analysis of TOI-1117\,b's photoevaporation history found rocky core and H/He atmosphere models to be inconsistent with observations, whilst water-rich scenarios were favoured.

\end{abstract}



\section{Introduction}
The Neptunian Desert is an area of period-radius and period-mass parameter space where significantly fewer exoplanets have been discovered. Initially identified by \citet{Szabo2011}, the boundaries of the desert were later defined by \citet{Mazeh2016} and recently refined by \citet{Deeg2023} and \citet{CastroGonzalez2024:neptunian-desert}. Recent studies have found a handful of exoplanets residing deep in the Neptunian Desert: TOI-332b \citep{Osborn2023}, NGTS-4b \citep{West2019}, LTT-9779b \citep{Jenkins2020}, TOI-849b \citep{Armstrong2020}, TOI-2196b \citep{Persson2022}, K2-100b \citep{Barragan2019}, K2-278b \citep{Livingston2018}, Kepler-644b \citep{Morton2016}, and TOI-1853b \citep{Naponiello2023}. Current theories explaining the lack of exoplanets in the Neptunian Desert include photoevaporation and tidal disruption, which induces strong tidal forces and extreme temperature variations, stripping the planets of their gas envelopes and leaving Earth-sized rocky cores \citep[e.g.][]{Matsakos2016,Owen2018}.

The Radius Valley is a similarly-unpopulated area of radius parameter space identified by \textit{Kepler} \citep{Borucki2010}. It describes a region of depressed planet occurrence in the distribution of planet radius and stellar irradiation, separating more commonly found rocky super Earths and sub-Neptunes with gas envelopes, along a line described approximately as $\log (R_p/R_\oplus) = 0.07\log (S/S_\oplus)+0.11$ \citep{Ho2023, Fulton2017}. Current planet evolution theories explaining the paucity of exoplanets in the Radius Valley include photoevaporation and core-powered mass loss \citep{Gupta2020}. Another theory is that the Radius Valley separates water-rich planets formed beyond the snow line, and water-poor planets formed inside the snow line \citep{Venturini2020}.

Neither the Neptunian Desert nor the Radius Valley can be attributed to observational biases as both Jupiter-sized and Earth-sized short-period planets were frequently detected by \textit{Kepler} transit observations and, since the detection bias is a monotonic function of planet radius at fixed orbital period, it cannot create a gap or feature at a specific radius within the observed range. Succeeding \textit{Kepler}, the Transiting Exoplanet Survey Satellite \textit{TESS} \citep{Ricker2015} searches for planets around bright hosts, which are optimal for detecting planetary radial velocity signals. After discovering a transiting planet, radial velocity measurements are conducted to obtain accurate planetary masses needed to identify planets located in the Neptunian Desert. With accurate masses and radii, densities can be determined, allowing internal structures to be analysed. Building a detailed population of Neptunian Desert planets could reveal the cause of the dearth. This is the purpose of the HARPS-NOMADS program, which has conducted the radial velocity follow-up observations for several \textit{TESS} objects-of-interest (TOIs), including TOI-1117.

This paper presents TOI-1117\,b, a transiting short-period sub-Neptune residing in the Neptunian desert, along with two outer sub-Neptunes, TOI-1117\,c and TOI-1117\,d, on longer, near-resonant orbits. Section~\ref{sec:Observations} presents photometric, spectroscopic, and speckle observations of TOI-1117. Section~\ref{sec:Results} provides the stellar and planetary parameters resulting from a joint fit of TESS, LCOGT, and HARPS observations. Section~\ref{sec:Analysis} describes various analyses conducted on the system architecture and interior structure of TOI-1117\,b. Section~\ref{sec:Discussion} discusses the implications on the potential formation and evolution mechanisms of the TOI-1117 system. Finally, Section~\ref{sec:Conclusions} concludes the paper.

\section{Observations}
\label{sec:Observations}
This section describes the observations of TOI-1117, which have been used to detect and characterise the planetary system. The system's astrometric and photometric properties are detailed in Table~\ref{table 1}. 
\begin{table}
    \caption{Details for TOI-1117.}
    \label{table 1}
    \begin{tabular}{llcl}
        \hline
        Property & (unit) & Value & Source\\
        \hline
        \\\multicolumn{2}{l}{\textbf{Identifiers}} \\
        TIC ID & & TIC 295541511 & TICv8.2\\
        TOI ID & & TOI-1117 & \textit{TESS}\\
        2MASS ID & & J18142452-6625113 & 2MASS\\
        \textit{Gaia} ID & & 6436995652638923392 & \textit{Gaia} DR2\\
        \\\multicolumn{2}{l}{\textbf{Astrometric properties}} \\
        RA & (J2000.0) & 18:14:24.49 & TICv8.2\\
        Dec. & (J2000.0) & -66:25:11.91 & TICv8.2\\
        Parallax & (mas) & $5.94\pm0.03$ & \textit{Gaia} DR2\\
        Distance & (pc) & $167.6\pm0.9$ & \textit{Gaia} DR2\\
        \\\multicolumn{2}{l}{\textbf{Photometric propeties}} \\
        \textit{TESS} & (mag) & $10.406\pm0.006$ & TICv8.2\\
        \it B & (mag) & $11.700\pm0.008$& TICv8.2\\
        \it V & (mag) & $11.016\pm0.023$& TICv8.2\\
        \it G & (mag) & $10.8621\pm0.0003$& TICv8.2\\
        \it J & (mag) & $9.770\pm0.022$& 2MASS\\
        \it H & (mag) & $9.474\pm0.027$& 2MASS\\
        \it K & (mag) & $9.385\pm0.022$& 2MASS\\
        \it W1 & (mag) & $9.64\pm0.022$& WISE\\
        \it W2 & (mag) & $9.409\pm0.021$& WISE\\
        \it W3 & (mag) & $9.336\pm0.033$& WISE\\
        \it W4 & (mag) & $8.93\pm0.42$& WISE\\
        \hline
        \multicolumn{4}{l}{\textit{Note. Sources:} TICv8.2 \citep{paegert2021}, 2MASS \citep{Skrutskie_2006}}\\
        \multicolumn{4}{l}{\textit{Gaia} Data Release 2 \citep{GAIA}, WISE \citep{Wright_2010}.}
    \end{tabular}
\end{table}

\subsection{TESS Photometry}
\label{sec:TESS}

This study uses \textit{TESS} observations of TOI-1117 from Sectors 13 and 39 with cadences of 30 minutes and 2 minutes, respectively. Following the Sector 13 observation, the candidate was alerted as a \textit{TESS} Object of Interest (TOI) by the identification of a transit signal from the MIT Quick-Look Pipeline \citep[QLP;][]{Huang_2020}. It was then re-observed in Sector 39 and both sets of data were processed by the \textit{TESS} Science Processing Operations Center \citep[SPOC;][]{2016SPIE.9913E..3EJ, TESSSPOC2020}, which is located at NASA Ames Research Center. The SPOC pipeline uses the Presearch Data Conditioning Simple Aperture Photometry \citep[PDCSAP;][]{Smith2012,Stumpe2012,Stumpe2014} to remove artificial trends generated by the instrument. No further detrending was required as the PDCSAP light curves appear flat across both sectors and no stellar activity was  seen in the HARPS data (see Section~\ref{sec:HARPS}). These data are available on the Mikulski Archive for Space Telescopes (MAST\footnote{\url{https://archive.stsci.edu/missions-and-data/transiting-exoplanet-survey-satellite-tess}}).

\begin{figure}
    \includegraphics[width=\columnwidth]{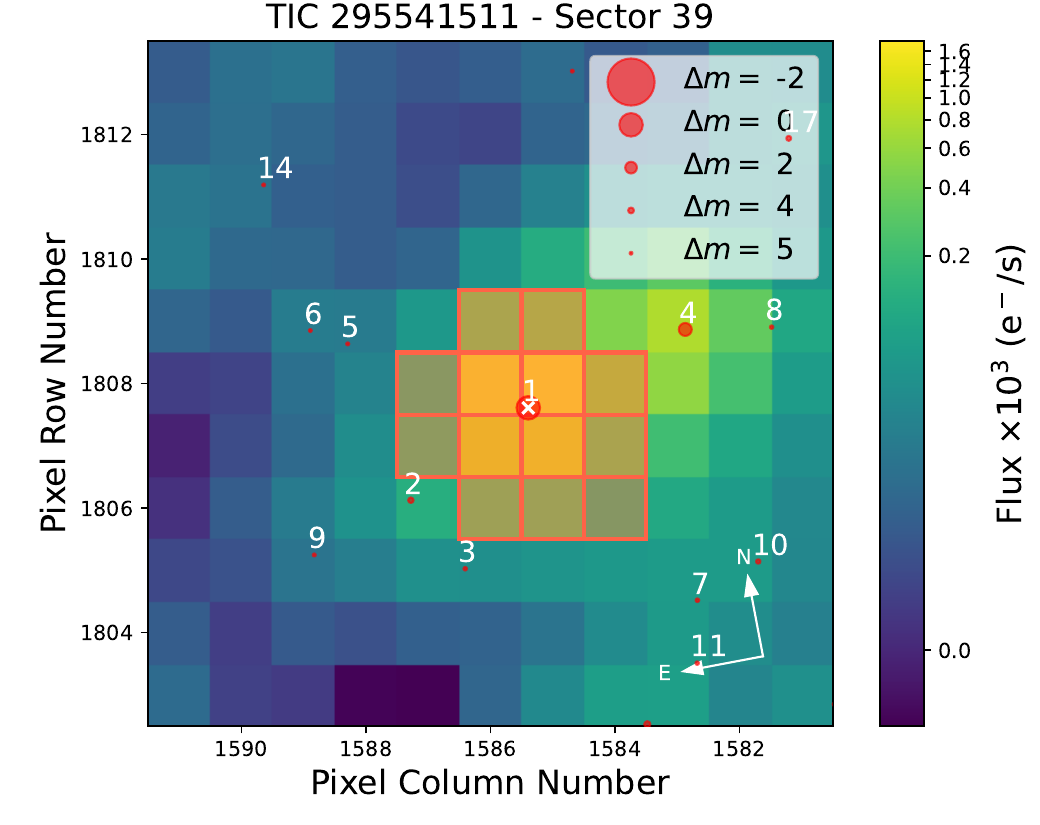}
    \caption{The average image across the time series in the Target Pixel File (TPF) for \textit{TESS} sector 39; showing TOI-1117 (white cross), the aperture mask used by SPOC (red squares), and other nearby \textit{Gaia} DR3 sources (red circles). Created using \texttt{TPFPLOTTER} \citep{Aller2020}}
    \label{fig:TPF}
\end{figure}

There are no nearby sources with magnitude contrasts $\rm \Delta m<5$ inside the SPOC aperture. However, there is a relatively bright source located 57.85 arcsec away from TOI-1117 (TIC 295541505, $G = 12.57$; Star 4 in Fig.~\ref{fig:TPF}) that could be contaminating the \textit{TESS} photometry. We used the \texttt{TESS-cont} algorithm\footnote{Availabe at \url{https://github.com/castro-gzlz/TESS-cont}} \citep{CastroGonzalez2024:TESS-cont} to quantify its flux contribution, finding a flux fraction of 1.6$\%$ in the aperture. This implies that a hypothetical eclipsing binary with a 3.7$\%$ transit depth in this star could have generated the observed eclipses, making it necessary to complement the \textit{TESS} data with ground-based observations of higher spatial resolutions. In Fig.~\ref{fig:tess-cont}, we plot the TESS contamination summary, where we can see the individual flux contributions from the five most contaminant sources. We note that the  PDCSAP photometry used in this work already accounts for flux dilution, so no additional corrections were necessary. In particular, accounting for all nearby sources, both the SPOC pipeline and \texttt{TESS-cont} estimate a CROWDSAP metric\footnote{The CROWDSAP metric indicates the target flux over the total flux falling inside the SPOC photometric aperture.} of 0.97, which was used by SPOC to correct the PDSCAP photometry. Furthermore, the difference image centroiding analysis, reported in the Sector 39 data validation report \citep{Twicken:DVdiagnostics2018}, located the host star within 2.9 +/- 2.6 arcsec of the transit source, complementing ground-based high resolution follow up. 

\begin{figure*}
    \includegraphics[width=\textwidth]{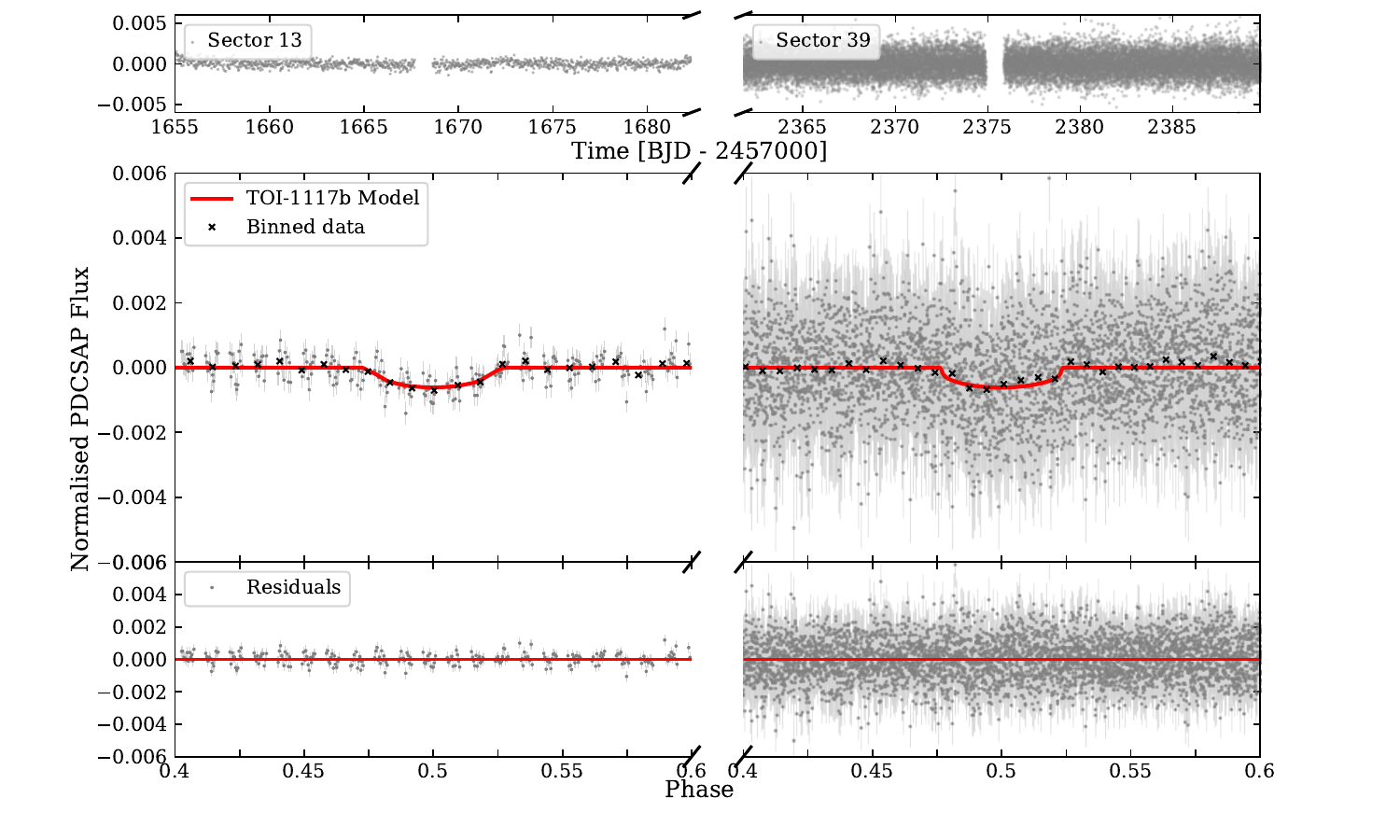}
    \caption{The top panel shows the TESS normalised TESS PDCSAP light curves for Sectors 13 (30-min cadence) and 39 (2-min cadence) as a time series in reduced Barycentric Julian Day (BJD). The middle panel shows the same data phase-folded with the joint fit model (Section~\ref{sec:jointfit}) plotted as a red line. The bottom panel shows the residuals of the joint fit.}    \label{fig:TESS}
\end{figure*}

Figure~\ref{fig:TESS} shows the normalised light curves and phase folded transits of TOI-1117\,b.

\subsection{Ground-based Photometry}
\label{sec:ground_based_phot}
To attempt to determine the true source of the \textit{TESS} detection, we acquired ground-based time-series follow-up photometry of the fields around TOI-1117 as part of the \textit{TESS} Follow-up Observing Program \citep[TFOP;][]{collins:2019}\footnote{\url{https://tess.mit.edu/followup}}. We used the {\tt TESS Transit Finder}, which is a customized version of the {\tt Tapir} software package \citep{Jensen:2013}, to schedule our transit observations. 

\subsubsection{M-Earth-South}
We observed a full transit window of TOI-1117\,b on 2019 August 11 from MEarth-South \citep{Irwin:2007}, which consisted of eight 0.4\,m telescopes located at Cerro Tololo Inter-American Observatory, east of La Serena, Chile. Each telescope used an Apogee U230 detector with a $29\arcmin\times29\arcmin$ field of view and an image scale of 0.84$\arcsec$ per pixel. Six of the seven telescopes were used to optimise target star photometry and the resulting light curve data were combined to attempt to detect the shallow event on-target. The seventh telescope overexposed the target star to improve the check for fainter nearby eclipsing binaries (NEBs). Results were extracted using the custom pipelines described in \cite{Irwin:2007}. We used circular photometric apertures with radius $5\farcs8$ that excluded all of the flux from the nearest known neighbor in the \textit{Gaia} DR3 catalog (\textit{Gaia} DR3 6437007399372647808), which is $\sim$17$\arcsec$ northwest of TOI-1117. The combination of the six on-target light curves had small systematic trends that caused the result to be inconclusive in the domain of the expected $\sim$600\,ppm transit depth.  Nearby star light curves were extracted from the seventh telescope observations and were checked for possible NEBs that could be contaminating the TESS photometric aperture. While all potentially contaminating stars near TOI-1117\,b were not conclusively ruled out as the source of the TESS detection, no obvious NEBs were found. This light curve was not used in the joint fit.

\subsubsection{LCOGT}
\begin{figure}
    \includegraphics[width=\columnwidth]{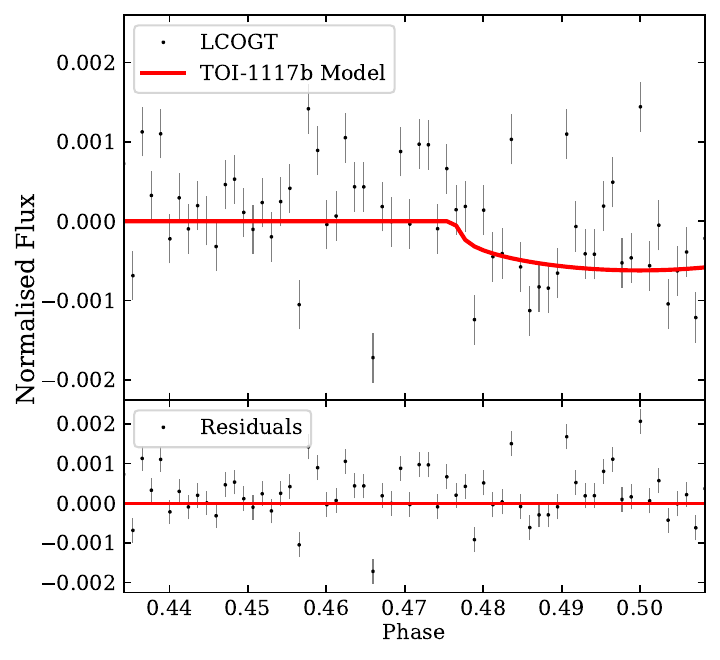}
    \caption{The top panel shows the detrended normalised LCO light curve with the results of the joint fit (Section~\ref{sec:jointfit}) plotted as an red line. The bottom panel shows the residuals of the joint fit.}
    \label{fig:LCO}
\end{figure}

We observed a partial transit window of TOI-1117\,b on 31 August 2019 from the Las Cumbres Observatory Global Telescope (LCOGT) \citep{Brown:2013} 1-m network node at South Africa Astronomical Observatory near Sutherland, South Africa (SAAO). The 1\,m telescope is equipped with a $4096\times4096$ pixel SINISTRO camera having an image scale of $0\farcs389$ per pixel, resulting in a $26\arcmin\times26\arcmin$ field of view and the images were calibrated by the standard LCOGT {\tt BANZAI} pipeline \citep{McCully:2018}, and differential photometric data were extracted using {\tt AstroImageJ} \citep{Collins:2017}. We used circular photometric apertures with radius $5\farcs8$ that excluded all of the flux from the nearest known neighbor in the \textit{Gaia} DR3 catalog. An approximately on-time, $\sim$600\,ppm ingress was detected on-target. Together, the LCOGT and M-Earth observations ruled out NEBs in all 109 neighboring stars that are bright enough to be capable of producing the TESS detection. The raw light curve data are available on the {\tt EXOFOP-TESS} website\footnote{\url{https://exofop.ipac.caltech.edu/tess/target.php?id=295541511}} and the LCOGT on-target lightcurve is included in the global modelling described in Section~\ref{sec:jointfit}. Figure~\ref{fig:LCO} shows the detrended normalised light curve, fitted ingress of TOI-1117\,b, and residuals. 

\subsection{HARPS Spectroscopy}
\label{sec:HARPS}

A total of 134 spectra were obtained via the HARPS-NOMADS program (PI: D. J. Armstrong, ID: 1108.C-0697) using the High Accuracy Radial velocity Planetary Searcher \citep[HARPS,][]{Mayor2003}. HARPS is a fibre-fed echelle spectrograph mounted on the ESO 3.6-m telescope at La Silla Observatory in Chile. The spectra were obtained between 6th June 2022 and 1st October 2023 using the high accuracy mode with an exposure time 1800s. The data were reduced with the standard offline HARPS data reduction pipeline and a G2 spectral mask was applied to the weighted cross correlation function (CCF) to determine the RV values.

\begin{figure}
   \includegraphics[width=\columnwidth,trim={0 1.5cm 0 2.5cm},clip]{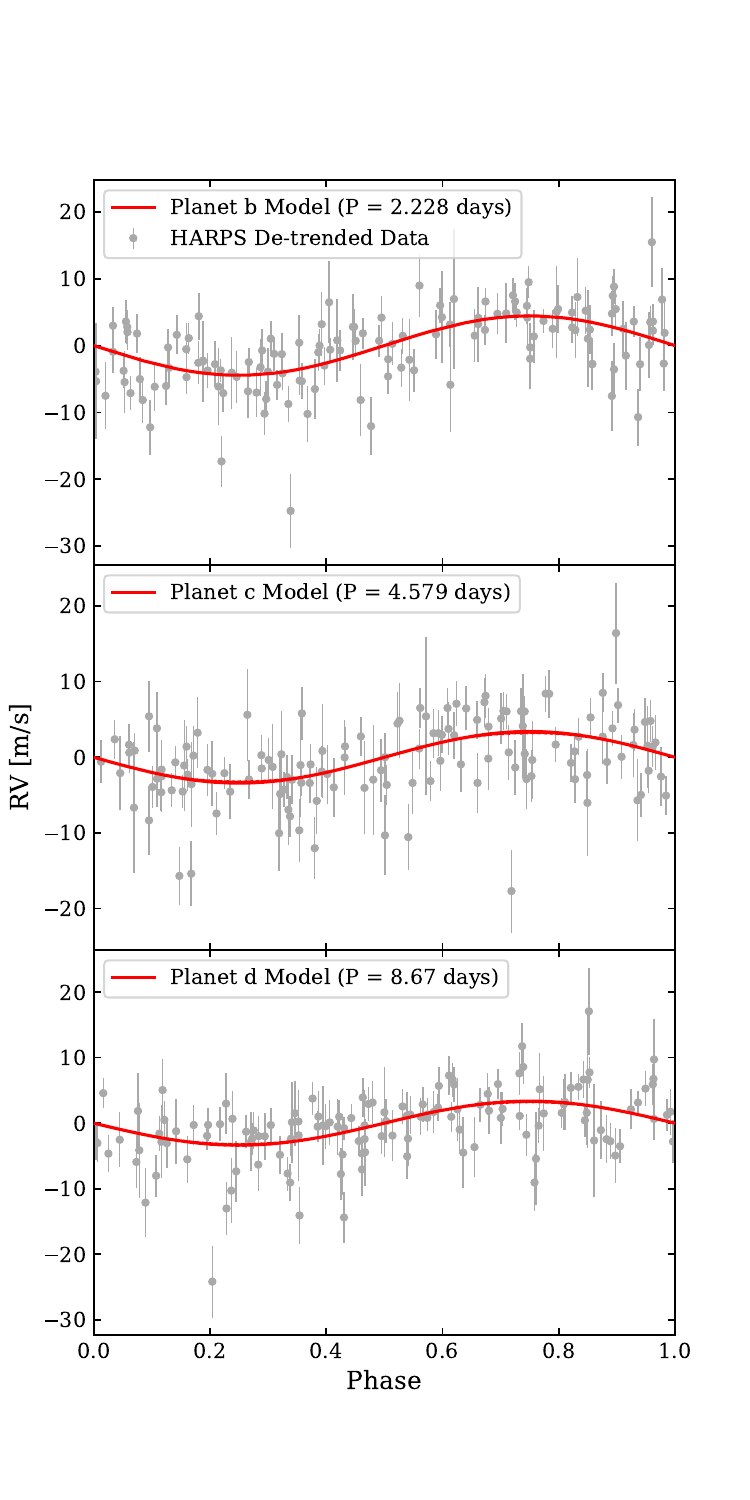} 
    \caption{Top: Phasefolded residuals of the HARPS data after removing the TOI-1117\,c and d signals, with the TOI-1117\,b planetary signal plotted as a red line. Middle: Phasefolded residuals of the TOI-1117\,b and d signals, with the TOI-1117\,c planetary signal plotted as a red line. Bottom: Phasefolded residuals of the TOI-1117\,b and c signals, with the TOI-1117\,d planetary signal plotted as a red line.}
    \label{fig:HARPS2}
\end{figure}

\begin{figure}
    \includegraphics[width=\columnwidth,trim={0cm 4cm 2cm 4cm},clip]{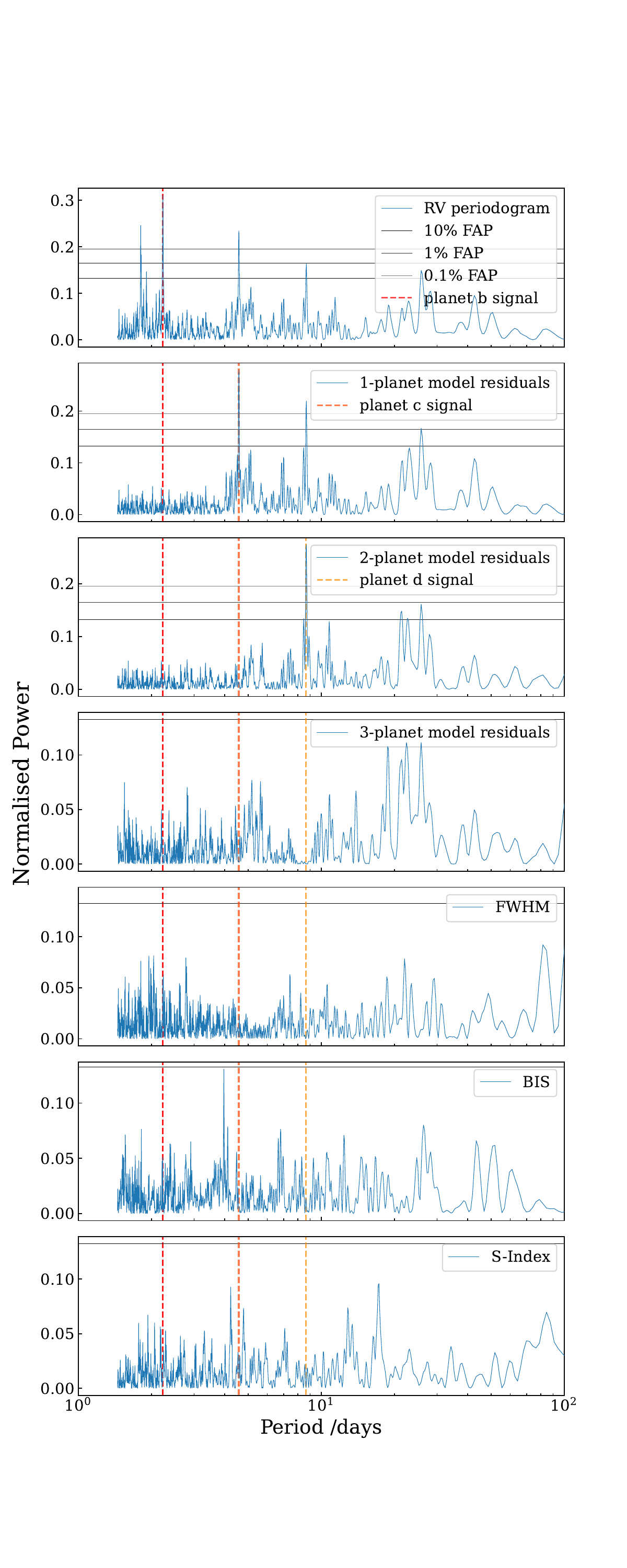}
    \caption{Lomb-Scargle Periodograms for the HARPS data, annotated with horizontal lines indicating the 10\%, 1\%, and 0.1\% false-alarm levels \citep{Baluev2008}. From top to bottom the periodograms show; the raw HARPS data with a significant peak at the expected orbital period of TOI-1117\,b (red); the residual from removing the first planetary signal, showing a significant peak at the expected period of TOI-1117\,c; the residuals of both planetary signals, showing a peak at the expected period of TOI-1117\,d; the residuals of all 3 planetary signals, showing no other significant periodic signal; the full-width-half-maximum of the raw HARPS data; the bisector span; and the S-index, showing no periodic signals in the stellar activity indicators.}
    \label{fig:GLS}
\end{figure}

Two data points were removed as they had anomalously high uncertainties (Table~\ref{tab:RVs}). The remaining data and the joint fit model can be seen in Figure~\ref{fig:HARPS2}. There is no periodicity or correlation between the radial velocities and the activity indicators (Table~\ref{tab:RVs}) and after removing the planetary signals, no other periodic signals are detected (Figure~\ref{fig:GLS}).

\begin{figure*}
    \includegraphics[width=\textwidth,trim={4cm 0 3cm 0},clip]{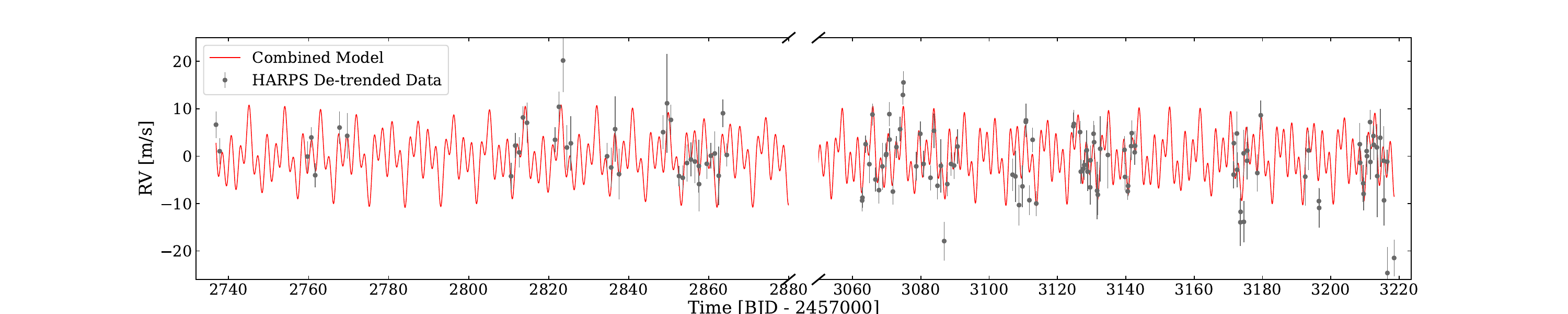}
    \caption{The HARPS radial velocity data (grey points), shown as a time series, with the combined 3-planet model from the joint fit (Section~\ref{sec:jointfit}) plotted as a red line.}   
    \label{fig:HARPS1}
\end{figure*}

\subsection{Speckle Imaging}
\label{sec:speckle}
High-angular resolution imaging is needed to search for nearby sources that can contaminate the \textit{TESS} photometry, which can lead to an underestimated planetary radius, or be the source of astrophysical false positives, such as background eclipsing binaries (EBs). If an exoplanet host star has a spatially-close companion, that companion (bound or line-of-sight) can create a false-positive transit signal. "Third-light” flux contamination from a close companion star will lead to an underestimated planetary radius if not accounted for in the transit model \citep[e.g.][]{Ciardi2015,CastroGonzalez2022} and cause non-detections of small planets residing within the same exoplanetary system \citep{Lester2021}. Additionally, the discovery of close, bound companion stars, which exist in nearly one-half of FGK type stars \citep{Matson2018}, provides crucial information toward our understanding of exoplanet formation, dynamics, and evolution \citep{Howell2021}. Thus, to search for EB companions unresolved in TESS or other ground-based follow-up observations, we obtained high-resolution imaging speckle observations of TOI-1117.

\subsubsection{SOAR Imaging}
\label{sec:SOAR}
\begin{figure}
    \includegraphics[width=\columnwidth]{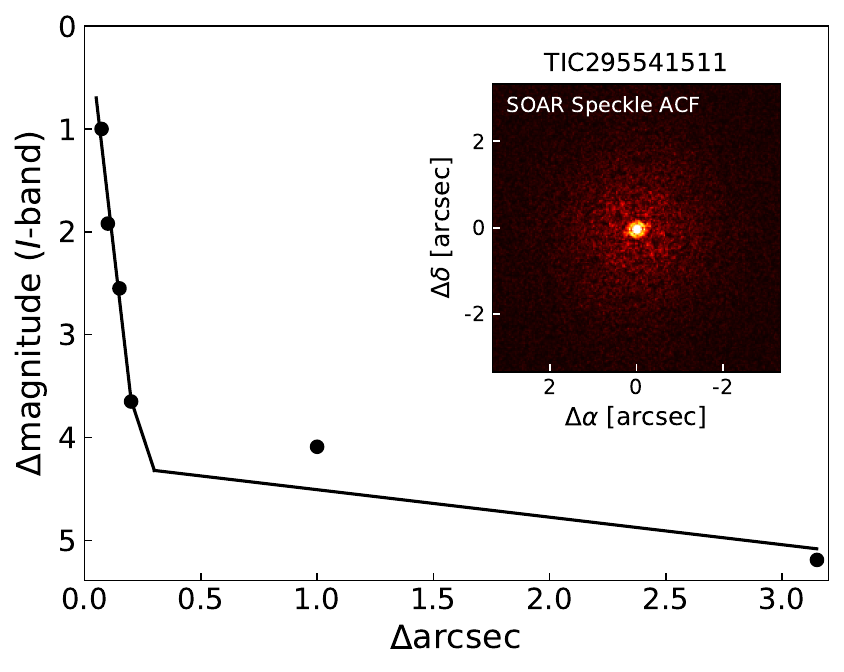}
 \caption{Plot showing the $5\sigma$ detection sensitivity and speckle auto-correlation functions (inset) from the observations in Cousins I-band as described in Section~\ref{sec:SOAR}. The star, TOI-1117 was found to have no bright close companions within $3\arcsec$.}
 \label{fig:speck}
\end{figure}

 We searched for stellar companions to TOI-1117 with the speckle imager on the 4.1m Southern Astrophysical Research (\textit{SOAR}) telescope \citep{Tokovinin2018} on 31 October 2020 UT, observing in Cousins I-band, a similar visible bandpass as \textit{TESS}. This observation was sensitive to companion stars 4.2 magnitudes fainter at an angular distance of $1\arcsec$ from the target. More details of the observations within the \textit{SOAR} \textit{TESS} survey are available in \cite{Ziegler2019}. The $5\sigma$ detection sensitivity and speckle auto-correlation functions from the observations are shown in Figure~\ref{fig:speck}. No nearby stars were detected within $3\arcsec$ of TOI-1117 in the SOAR observations.

\subsubsection{Gemini-8m Imaging}
\label{sec:Gemini}
\begin{figure}
    \includegraphics[width=\columnwidth]{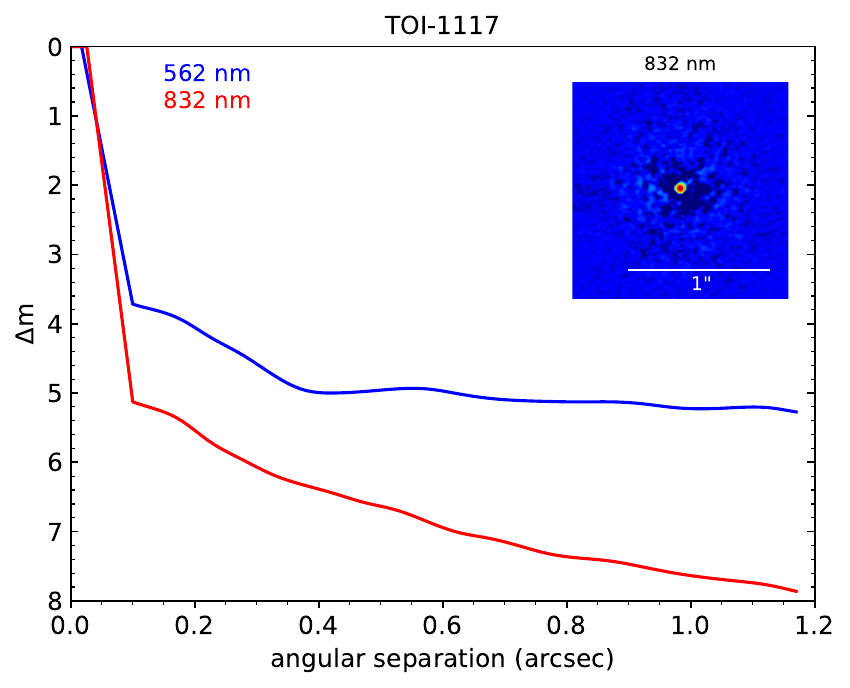}
    \caption{Plot showing the $5\sigma$ speckle imaging contrast curves in both filters as a function of the angular separation out to $1.2\arcsec$. The inset shows the reconstructed 832nm image with a $1\arcsec$ scale bar. The star, TOI-1117, was found to have no close companions from the diffraction limit out to $1.2\arcsec$to within the contrast levels achieved.}
    \label{fig:Gemini}
\end{figure}

TOI-1117 was also observed on 27 May 2023 UT using the Zorro speckle instrument on the Gemini South 8-m telescope \citep{Scott2021}.  Zorro provides simultaneous speckle imaging in two bands (562nm and 832nm) with output data products including a reconstructed image with robust contrast limits on companion detections. Five sets of $1000\times0.06$ sec exposures were collected and reduced via Fourier analysis with the speckle data reduction pipeline from \citet{Howell2011}. Figure~\ref{fig:Gemini} shows the final contrast curves and 832nm reconstructed speckle image. We find that TOI-1117 is a single star with no companion within 5-8 magnitudes of the target star from the diffraction limit of the telescope (20 mas) out to 1.2$\arcsec$. At the distance of TOI-1117 (d = 167.6 pc), these angular limits correspond to spatial limits of 3.4 to 201 au.

\section{Results}
\label{sec:Results}

\subsection{Stellar Analysis}
\label{sec:stellar}

We used the ARES+MOOG methodology to estimate the stellar spectroscopic parameters ($T_{\mathrm{eff}}$, $\log g$, microturbulence, [Fe/H]). This methodology is described in detail in \citet[][]{Sousa-21, Sousa-14, Santos-13}. The equivalent widths (EW) were consistently measured using the ARES code\footnote{The last version, ARES v2, can be downloaded at \url{https://github.com/sousasag/ARES}} \citep{Sousa-07, Sousa-15}. For the spectral analysis we used the list of iron lines originally presented in \citet[][]{Sousa-08} which were analysed in a combined HARPS spectrum for TOI-1117. To converge for the best set of spectroscopic parameters for each spectrum we use a minimization process to find the ionization and excitation equilibrium. This process makes use of a grid of Kurucz model atmospheres \citep{Kurucz-93} and the latest version of the radiative transfer code MOOG \citep{Sneden-73}. We also derived a more accurate trigonometric surface gravity using recent \textit{Gaia} data following the same procedure as described in \citet[][]{Sousa-21} which provided a consistent value when compared with the spectroscopic surface gravity. In this process we also estimate the stellar mass using the calibration presented in \citet[][]{Torres-2010}. The stellar radius was also derived using a similar calibration presented in in \citet[][]{Torres-2010}.

Using the aforementioned stellar atmospheric parameters and MOOG, we determined the abundances of refractory elements following the classical curve-of-growth analysis method described in \citep[e.g.][]{Adibekyan-12, Adibekyan-15, DelgadoMena-17}. Similar to the stellar parameter determination, we used ARES to measure the EWs of the spectral lines of these elements, and used a grid of Kurucz model atmospheres \citep{Kurucz-93} along with the radiative transfer code MOOG \citep{Sneden-73} to convert the EWs into abundances, assuming local thermodynamic equilibrium. The abundances of the volatile elements C and O were derived with the same tools following the methodology by \citep{DelgadoMena-21} and \citep{Bertrandelis-15}. The oxygen abundances are based on two weak atomic lines. One of them, located at 6300\AA{}, is often contaminated by an Earth airglow and hence we could only use half of the HARPS spectra to determine its abundance, lowering the S/N of the combined spectrum. Therefore, the error in oxygen abundance is larger than for the other elements. We obtained all the [X/H] ratios by doing a differential analysis with respect to a high S/N solar (Vesta) spectrum from HARPS (see Table \ref{table A1}).

In addition, we obtained the abundance of lithium by performing spectral synthesis with MOOG, following the same procedure as in \citet{DelgadoMena-14}. We first fixed the macroturbulence velocity to 3.1 km\,s$^{-1}$ \citep[based on the empirical calibration by][]{doyle-14} to estimate the $v \, \textrm{sin} \, i_{\star}$ from two Fe lines in the region, leading to a value of 1.5 $\pm$ 0.1 km\,s$^{-1}$. We obtained an abundance A(Li)\,<\,0.3\,dex, which is an expected value for a not young star of this $T_{\rm eff}$. 

Finally, we used the chemical abundances of some elements to derive ages through the so-called chemical clocks (i.e., certain chemical abundance ratios that have a strong correlation with age). We applied the 3D formulas described in Table 10 of \citet{DelgadoMena-19}, which also consider the variation in age produced by the effective temperature and iron abundance. The chemical clocks [Y/Mg], [Y/Zn], [Y/Ti], [Y/Si], [Sr/Mg], [Sr/Zn], [Sr/Ti], and [Sr/Si] were used from which we obtain a weighted average age of 4.42\,$\pm$\,0.85 Gyr. We note that this is a weighted average error from the individual clocks and thus reflects the good agreement between the different abundance ratios. We will consider as final error for this method a more conservative value of 1.5\,Gyr since this is the maximum error in age of the stars employed to derive the aforementioned calibrations. An age estimate was also obtained by using \texttt{pyrhk} \citep{Silva2019} to calculate $logR'_{HK}=-5.03\pm0.02$, and then applying gyrchronology relations to estimate a rotation period of approximately $\sim$40 days and an age $\sim$7 Gyr \citep{Mamajek2008}. We chose to use the chemical clocks results because the gyrochronology results are less precise.

\begin{table}
 \caption{Stellar parameters of the TOI-1117 system.}
 \label{table 2}
 \centering
  \begin{tabular}{llr}
   \hline
   Parameter & Unit & Value \\
   \hline\\
   \multicolumn{2}{l}{\textbf{Bulk parameter}}\\
   $T_{\rm eff}$& K & $5635\pm 62$\\
   log $g_*$ (spec) & cm\,s$^{-2}$ & $4.35\pm 0.11$\\
   log $g_*$ (\textit{Gaia}) & cm\,s$^{-2}$ & $4.41\pm 0.03$\\
   $v_\mathrm{mic}$ & km~s$^{-1}$        & $0.93  \pm 0.02$\\ 
   $[$Fe/H$]$ & & $0.136\pm0.044$\\
   Spectral type & & G5V\\
   Mass & $M_{\odot}$ & $0.97\pm0.02$\\
   Radius & $R_{\odot}$ & $1.05\pm0.03$\\
   $\rho_*$ & g\,cm$^{-3}$ & $1.28\pm0.18$\\
   Luminosity & $L_\odot$ & $0.36\pm0.01$\\
   Age & Gyr & $4.42\pm1.50$\\
   $\log{R'_\text{HK}}$ & & $-5.03\pm0.02$\\
   $P_\text{rot}$ & days & $\sim40$\\
   $v \sin\textit{i}$ & km~s$^{-1}$ & $1.32\pm0.04$\\
   \hline
 \end{tabular}
\end{table}

\subsection{Joint Fit}
\label{sec:jointfit}

The TESS and LCOGT photometry and HARPS RVs were combined into a joint fit by using python packages \texttt{exoplanet} \citep{F-M2021} and \texttt{celerite2} \citep{celerite2}, which employ \texttt{Starry} \citep{starry} to create transit models and \texttt{PyMC3} \citep{pymc3} to draw samples from posteriors using Monte Carlo methods. \texttt{exoplanet} adopts a Keplerian orbit model parameterised by the stellar mass $M_\ast$ and radius $R_\ast$, the orbital period $P$, and time of midtransit $T_0$, the impact parameter $b$, the eccentricity $e$, and the argument of periastron $\omega$. Gaussian prior distributions for each parameter were input into the model, centered around values either found by stellar analysis (Section~\ref{sec:stellar}) or estimated by DACE\footnote{https://dace.unige.ch}. Also included in the fit was a quadratic limb-darkening parameter parametrised by \citet{Kipping} and a jitter parameter, centered around the minimum RV uncertainty, to account for instrumental effects and short-scale stellar activity.

We did not include any larger-scale stellar activity parameters because the Lomb-Scargle periodograms (Figure~\ref{fig:GLS}) show no significant (<10\% FAP) signals in the stellar activity indicators. Furthermore, the jitter parameter results in 4.5m/s and the median HARPS error is 3.2 m/s so the average combined error is 5.5 m/s. 95\% of the HARPS measurements fall within 1$\sigma$ and all are within 3$\sigma$. Therefore, there is no evidence of periodic radial velocity changes due to stellar activity.  

The periodograms do, however, show three significant (<0.1\% FAP) periodic signals in the RVs, therefore we fitted for three Keplerian orbits. To further justify the decision to adopt a three-planet model, we tested one-planet, two-planet, and four-planet models using similar \texttt{PyMC3} setups and the same sampling configuration as the three-planet model. The potential fourth planet was given a uniform prior for the period, covering the whole time span of the photometric datasets. In order to evaluate model complexity and predictive performance, we computed the Widely Applicable Information Criterion (WAIC; \citet{watanabe2010}). WAIC is defined as $ -2\, \times$ expected log pointwise predictive density (ELPD), so a lower WAIC indicates better predictive performance. The WAIC for the photometric datasets showed no significant changes, with the standard errors of the WAIC values making $\Delta$WAIC consistent with 0. This is expected as only one planet is seen to transit. However, for the HARPS spectroscopy, the one-planet and two-planet models yield higher WAIC values, showing strong evidence for worse predictive performances ($\Delta\text{WAIC}\gg10$). The four-planet model shows no significant improvement in fitting to the HARPS data ($\Delta\text{WAIC}\ll10$). We therefore select the 3-planet model as the most appropriate representation of the observed data based on predictive accuracy.

The joint fit was initially conducted without a constraint on the eccentricities, $e$, or argument of periastron, $\omega$, of the three planets, using {\tt pymc3-ext.UnitDisk} on [$\sqrt{e}\sin{\omega}, \sqrt{e}\cos{\omega}$] to generate a prior uniform distribution on a two dimensional unit disk. This "eccentric joint fit" found:\begin{center} $e_b=0.05^{+0.06}_{-0.04}$\quad $e_c=0.13^{+0.16}_{-0.09} $\quad $e_d=0.11^{+0.13}_{-0.08}$\\[2pt] $\omega_b=-1.2^{+2.3}_{-0.9}$\quad $\omega_c
=0.9^{+2.4}_{-1.1}$\quad $\omega_d=-0.01^{+2.08}_{-1.89}$.\\\end{center}  The eccentricities of all three planets were consistent with 0 and the arguments of periastron was poorly constrained. Therefore, all $e$ and $\omega$ were fixed to 0 for the final joint fit. This assumption is further justified by stability analysis (Section~\ref{sec:stability}). 

The prior distributions of the final joint fit are listed in Table~\ref{tab:priors} and initial fit values were found by maximising the log probability from \texttt{PYMC3}. These posteriors were then used to draw 300,000 samples (25 chains, 2,000 discarded as tuning per chain). The best fit results are displayed in Table~\ref{tab:planets}.

\begin{table*}
 \caption{TOI-1117 system and instrumental parameters input to \texttt{exoplanet}.}
 \label{tab:priors}
  \begin{tabular}{lllr}
   \hline
   Parameter & unit & Prior Distribution* & Fit Value \\
   \hline
   \\{\textbf{Photometry}}\\
   Mean (Sector 13) & - & $\mathcal{N}\left(0,1\right)$ & $2.13\pm1.14\times10^{-5}$\\
   Mean (Sector 39) & - & $\mathcal{N}\left(0,1\right)$ & $4.57\pm0.97\times10^{-5}$\\
   Mean (LCOGT) & - & $\mathcal{N}\left(0,1\right)$ & $8.38\pm2.57\times10^{-5}$\\
   Limb-darkening coefficient $u_1$**& - & \cite{Kipping} & $0.61^{+0.25}_{-0.26}$\\
   Limb-darkening coefficient $u_2$& - & \cite{Kipping} & $0..64^{+0.24}_{-0.32}$\\
   \\{\textbf{HARPS Spectroscopy}}\\
   Offset & ms$^{-1}$ & $\mathcal{U}\left(-24900.0,-24890\right)$ & $-24894.386^{+0.347}_{-0.354}$\\
   ln(Jitter) & ms$^{-1}$ & $\mathcal{U}\left(0,5\right)$ & $1.50^{+0.41}_{-0.53}$\\
   \\{\textbf{TOI-1117}}\\
   Mass $M_{\ast}$ & $M_{\odot}$ & $\mathcal{N}_{\mathcal{B}}\left(0.97,0.02,0.0,3.0\right)$ & $0.97\pm0.02$\\
   Radius $R_{\ast}$ & $R_{\odot}$ & $\mathcal{N}_{\mathcal{B}}\left(1.05,0.03,0.0,3.0\right)$ & $1.06\pm0.03$\\
  \\ {\textbf{TOI-1117\,b}}\\
   Orbital Period \textit{P} & days & $\mathcal{U}\left(2,2.5\right)$ & Table~\ref{tab:planets}\\
   RV Semi-Amplitude \textit{K} & ms$^{-1}$  & $\mathcal{U}\left(0,50\right)$ & Table~\ref{tab:planets}\\
   Midtransit Time $T_0$& BJD-2457000 & $\mathcal{U}\left(2382,2390\right)$ & Table~\ref{tab:planets}\\
   Eccentricity \textit{e} & - & Fixed & 0\\
   Argument of Periastron $\omega$ & deg & Fixed & 0\\
   log(Radius) & $R_{\odot}$ & $\mathcal{N}\left(-3.59062,1\right)$ & $-3.79\pm0.05$\\
   Impact Parameter \textit{b} & - & $\mathcal{U}\left(0,1.0228\right)$ & $0.17^{+0.15}_{-0.12}$\\
   \\{\textbf{TOI-1117\,c}} \\
   Orbital Period \textit{P} & days & $\mathcal{U}\left(4,5\right)$ & Table~\ref{tab:planets}\\
   RV Semi-Amplitude \textit{K} & ms$^{-1}$  & $\mathcal{U}\left(0,50\right)$ & Table~\ref{tab:planets}\\
   Midtransit Time $T_0$& BJD-2457000 & $\mathcal{U}\left(2490,2505\right)$ & Table~\ref{tab:planets}\\
   Eccentricity \textit{e} & - & Fixed & 0\\
   Argument of Periastron $\omega$ & deg & Fixed & 0\\
   \\{\textbf{TOI-1117\,d}}\\
   Orbital Period \textit{P} & days & $\mathcal{U}\left(8,9\right)$ & Table~\ref{tab:planets}\\
   RV Semi-Amplitude \textit{K} & ms$^{-1}$  & $\mathcal{U}\left(0,50\right)$ & Table~\ref{tab:planets}\\
   Midtransit Time $T_0$& BJD-2457000 & $\mathcal{U}\left(2485,2505\right)$ & Table~\ref{tab:planets}\\
   Eccentricity \textit{e} & - & Fixed & 0\\
   Argument of Periastron $\omega$ & deg & Fixed & 0\\
   \hline
   \multicolumn{4}{l}{\footnotesize{*$\mathcal{N}\left(\mu,\sigma\right)$ denotes a normal distribution with a mean $\mu$ and a standard deviation $\sigma$.}}\\
   \multicolumn{4}{l}{\footnotesize{$\mathcal{N}_{\mathcal{B}}\left(\mu,\sigma,a,b\right)$ denotes a bounded normal distribution with lower bound a and upper bound b.}}\\
   \multicolumn{4}{l}{\footnotesize{$\mathcal{U}\left(a,b\right)$ denotes a uniform distribution with lower bound a and upper bound b.}}\\
   \multicolumn{4}{l}{\footnotesize{**The distributions for limb-darkening coefficients are built into the \texttt{exoplanet} framework.}}
 \end{tabular}
\end{table*}

\begin{table}
 \caption{Planetary parameters for the TOI-1117 system from \texttt{exoplanet}.}
 \label{tab:planets}
  \begin{tabular}{llr}
   \hline
   Parameter & unit & Value \\
   \hline\\
   {\textbf{TOI-1117\,b}}\\
   \multicolumn{3}{l}{-------------------------------- Fitted Parameters ---------------------------------}\\[2pt]
   Orbital Period \textit{P} & days & $2.22816\pm0.00001$\\
   RV Semi-Amplitude \textit{K} & ms$^{-1}$  & $4.45\pm0.47$\\
   Midtransit Time $T_0$& BJD-2457000 & $2386.964\pm0.004$\\
   Eccentricity \textit{e} & - & 0\\
   \multicolumn{3}{l}{------------------------------- Derived Parameters -------------------------------}\\[2pt]
   Mass $M_\text{p}$ & $M_\oplus$ & $8.90^{+0.95}_{-0.96}$\\[2pt]
   Radius $R_\text{p}$ & $R_\oplus$ & $2.46^{+0.13}_{-0.12}$\\
   Bulk Density $\rho$ & g\,cm$^{-3}$ & $3.30\pm0.60$\\
   Semi-Major Axis \textit{a} & AU & $0.0330\pm0.0002$\\
   Inclination \textit{i} & rad & $1.54\pm{0.02}$\\
   Longitude of Ascending Node $\Omega$ & rad & $-0.003\pm{2.135}$\\
   Impact Parameter \textit{b} & - & $0.17^{+0.15}_{-0.12}$\\
   Equilibrium Temperature $T_\text{eq}$* & K & $1538\pm21$\\
   Insolation Flux \textit{S} & $S_\oplus$ & $928\pm50$\\
   \\{\textbf{TOI-1117\,c}} \\
   \multicolumn{3}{l}{-------------------------------- Fitted Parameters ---------------------------------}\\[2pt]
   Orbital Period \textit{P} & days & $4.579\pm0.004$\\
   RV Semi-Amplitude \textit{K} & ms$^{-1}$ & $3.45\pm0.47$\\
   Eccentricity \textit{e} & - & 0\\
   \multicolumn{3}{l}{------------------------------- Derived Parameters -------------------------------}\\[2pt]
   Minimum Mass $M_\text{p}\sin{i}$ & $M_\oplus$ & $8.78^{+1.19}_{-1.21}$\\
   Mass (See Section~\ref{sec:stabilityinc}) & $M_\oplus$ & $7.57<M_c<11.3$\\
   Semi-Major Axis \textit{a} & AU & $0.0534\pm0.0003$\\
   Equilibrium Temperature $T_\text{eq}$* & K & $1210\pm16$\\
   Insolation Flux \textit{S} & $S_\oplus$ & $355\pm19$\\
   \\{\textbf{TOI-1117\,d}}\\
   \multicolumn{3}{l}{-------------------------------- Fitted Parameters ---------------------------------}\\[2pt]
   Orbital Period \textit{P} & days & $8.665\pm0.011$\\
   RV Semi-Amplitude \textit{K} & ms$^{-1}$ & $3.40\pm0.50$\\
   Eccentricity \textit{e} & - & 0\\
   \multicolumn{3}{l}{------------------------------- Derived Parameters -------------------------------}\\[2pt]
   Minimum Mass $M_\text{p}\sin{i}$ & $M_\oplus$ & $10.71^{+1.59}_{-1.58}$\\
   Mass (See Section~\ref{sec:stabilityinc}) & $M_\oplus$ & $9.19<M_c<14.2$\\
   Semi-Major Axis \textit{a} & AU & $0.0817\pm0.0006$\\
   Equilibrium Temperature $T_\text{eq}$* & K & $978\pm13$\\[2pt]
   Insolation Flux \textit{S} & $S_\oplus$ & $152\pm8$\\
   \hline
   \multicolumn{2}{l}{\footnotesize{*Assuming albedo=0}}
 \end{tabular}
\end{table}

\begin{table}
    \caption{Differences in Widely Applicable Information Criterion (WAIC) values for 1-planet, 2-planet, and 4-planet joint fit models, computed relative to the 3-planet model used in our analysis. WAIC is defined as $ -2\, \times$ expected log pointwise predictive density (ELPD), so a lower WAIC indicates better predictive performance.}
    \label{tab:WAIC}
    \begin{tabular}{lcccc}
    \hline
    \multicolumn{1}{}{} & \multicolumn{4}{c}{$\Delta$ WAIC (relative to 3-planet model)}\\
    Model & HARPS & TESS S39 & TESS S13 & LCOGT\\
    \hline\\
    1-planet     & $63.00$ & $0.02$ & $0.06$ & $0.02$\\
    2-planet    & $36.44$ & $-0.04$ & $0.04$ & $0.02$\\
    3-planet     &  0 & 0 & 0 & 0 \\
    4-planet    & $-2.48$ & $-0.4$ & $-0.36$ & $0.02$\\
    \end{tabular}
\end{table}

\begin{figure}
    \includegraphics[width=\columnwidth]{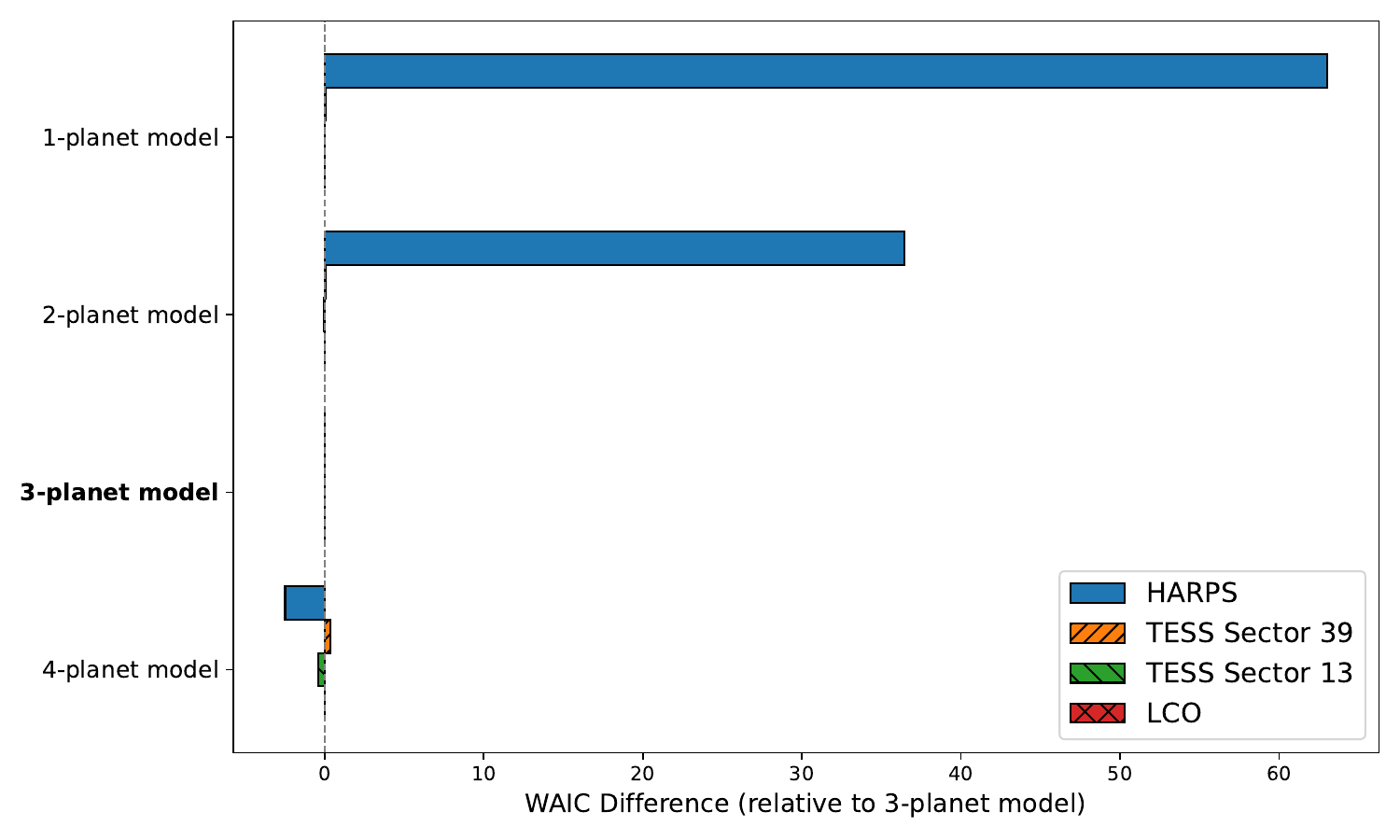}
    \caption{Bar plot of differences in Widely Applicable Information Criterion (WAIC) values for 1-planet, 2-planet, and 4-planet joint fit models, computed relative to the 3-planet model used throughout this work.}
    \label{fig:WAIC}
\end{figure}

\section{Analysis}
\label{sec:Analysis}

\subsection{Stability Analysis}
\label{sec:stability}

\subsubsection{Eccentricity}
\label{sec:stabilityecc}
The aim of this stability analysis is to constrain a range of eccentricities for which the system is stable, with the aim of testing the decision to set the eccentricity of all three planets to 0 in the final joint fit (see Section ~\ref{sec:jointfit}). 

An eccentric joint fit was conducted without a constraint on the eccentricities or argument of periastron of the three planets, using {\tt pymc3-ext.UnitDisk} on [$\sqrt{e}\sin{\omega}, \sqrt{e}\cos{\omega}$] to generate a prior uniform distribution on a two dimensional unit disk. From 500 randomly selected posterior samples, we simulated various potential architectures for the TOI-1117 system using \texttt{rebound} \citep{Rein2012}. \texttt{spock} \citep{Tamayo2020} was then used to calculate the median predicted time after which the system would become unstable. In this case, a system with median instability time longer than 1Myr was considered stable.

\begin{figure}
    \includegraphics[width=\columnwidth,trim={0 0 0 1cm},clip]{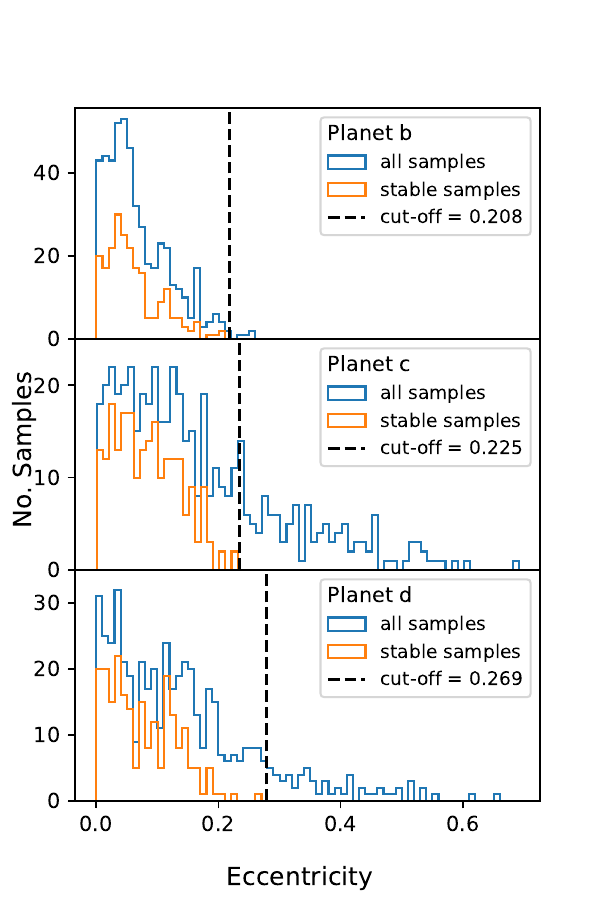}
     \caption{The distribution of eccentricities for TOI-1117\,b, c, and d for 500 simulated orbits (blue histogram) and those of which have a predicted instability time longer than 1Myr (orange histogram). The black dashed lines indicate the maximum eccentricities where the system is stable. Examples of simulated system architectures can be seen in Figure~\ref{fig:Orbits}}
    \label{fig:stabilityecc}
\end{figure}

We found only 45\% of samples to have a median instability time $>$\,1\,Myr with the maximum stable eccentricities being $e_b=0.208$, $e_c=0.225$ and $e_d=0.269$. Figure~\ref{fig:stabilityecc} shows the eccentricity distributions of the stable simulations. Furthermore, no simulations with non-zero eccentricities have an expected stability time $>$\,1\,Gyr, whilst all simulations with eccentricities of 0 were stable after 1\,Gyr. With the age of TOI-1117 being $\gg$1\,Gyr, this supports the decision to set the eccentricity of all three planets to 0 in the final joint fit. Example system architectures are shown in Figure~\ref{fig:Orbits}.



\subsubsection{Inclination}
\label{sec:stabilityinc}

\begin{figure}
    \includegraphics[width=\columnwidth,trim={0 0 1cm 0},clip]{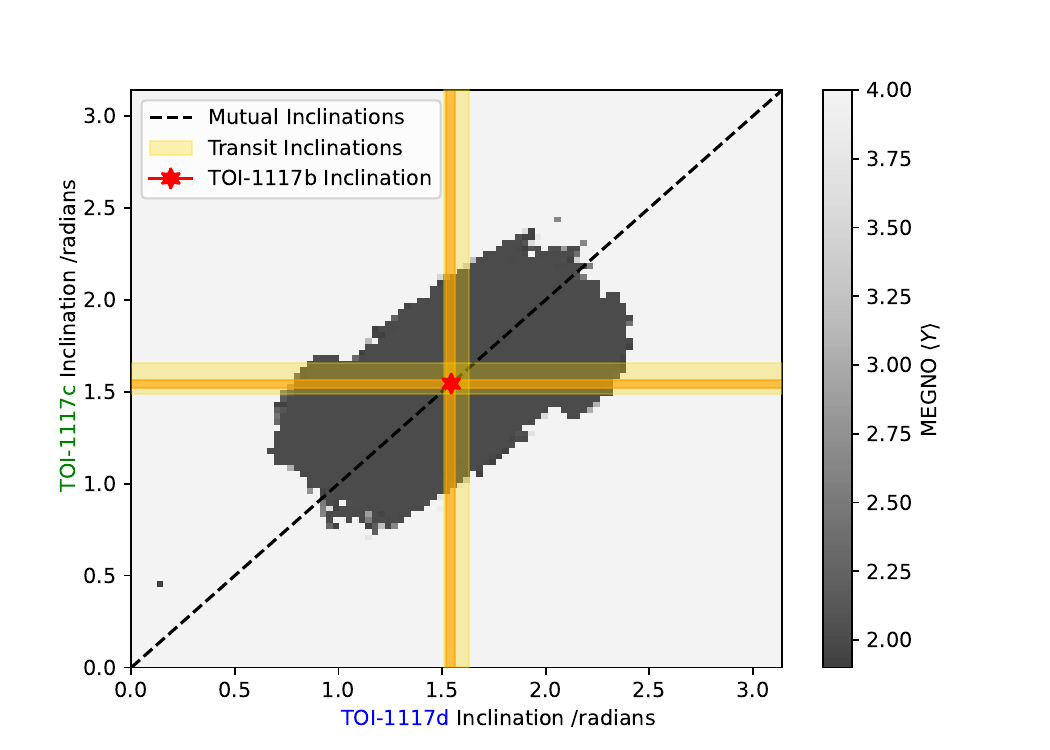}
    \caption{The distribution of the MEGNO chaos indicator for simulated TOI-1117 systems with a full range of possible inclinations for TOI-1117\,c and d. A MEGNO value of 2 corresponds to regular quasi-periodic orbits and higher values indicate chaotic orbits. Other system parameters are taken from the results of the joint model (Table~\ref{tab:planets}) and the longitude of the ascending node is set as 0 for all 3 planets. Examples of simulated system architectures can be seen in Figure~\ref{fig:inclinedOrbits}.}
    \label{fig:MEGNO}
\end{figure}

Another use of this method is to assess the impact of the unknown inclinations and therefore poorly constrained masses for planet c and d. By simulating the system with varying inclinations of TOI-1117\,c and d, a more precise range of possible masses can be found. A MEGNO (Mean Exponential Growth of Nearby Orbits) value is calculated by \texttt{rebound} for each combination of inclinations. This MEGNO value is a chaos indicator, defined by \cite{Cincotta2000} and shown to be effective by \cite{Maffione2011}. A MEGNO value of 2 corresponds to regular quasi-periodic orbits and higher values indicate chaotic orbits. Figure~\ref{fig:MEGNO} shows the results of this inclination analysis, displaying a preference to co-orbital system architectures and ruling out the possibilities of inclinations outside of the range $1.14<i_c<2.06$ and $0.70<i_d<2.44$ radians, relative to the line of sight, and therefore providing upper limits to the masses $M_c<11.3M_\oplus$ and $M_d<14.2M_\oplus$.

\subsection{Resonance Analysis}
\label{sec:resonance}

\begin{figure}
    \includegraphics[width=0.9\columnwidth]{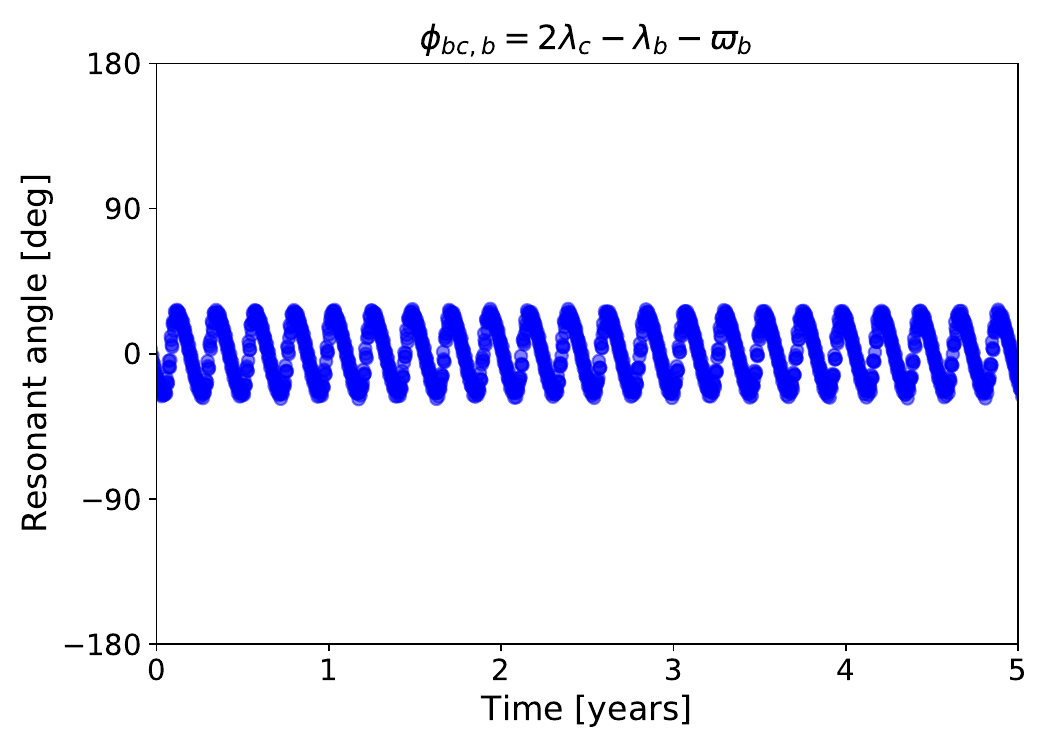}
    \includegraphics[width=0.9\columnwidth]{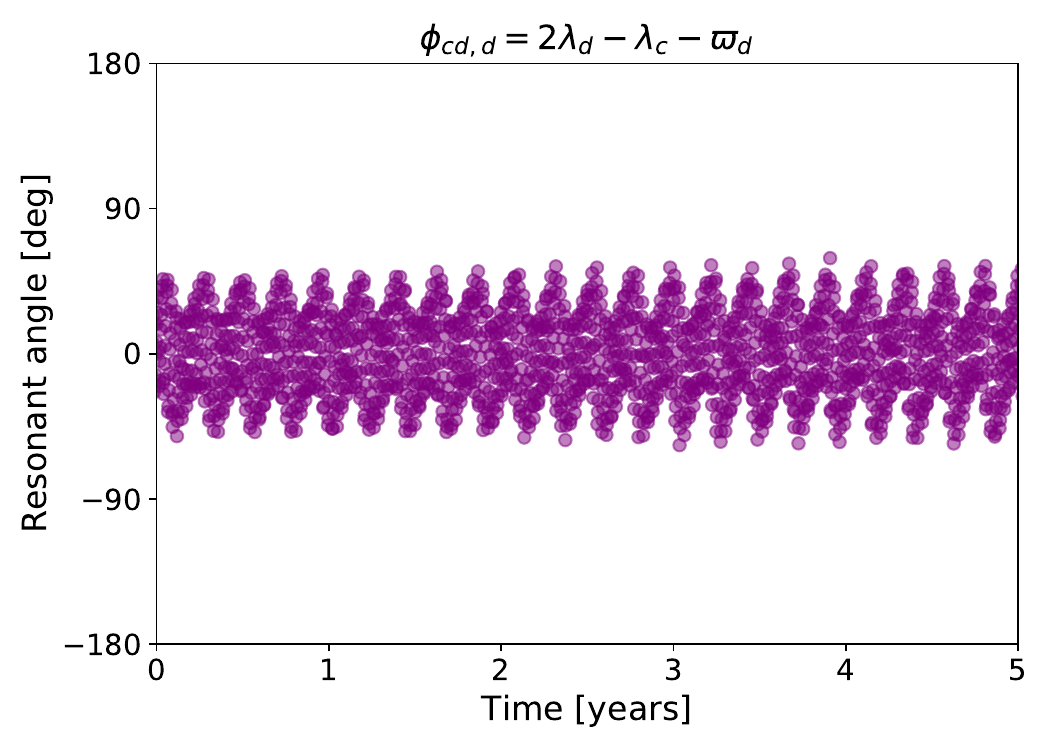}
    \caption{Examples of the time evolution of the critical resonant angles for two $N$-body integrations that showed resonant trajectories. The top and bottom panels show the evolution of one of the resonant angles for the b/c pair and c/d pair, respectively.}
    \label{fig:critical angles}
\end{figure}

The TOI-1117 planets appear close to mean-motion resonances (MMRs). The period ratios of the adjacent pairs are $P_c/P_b\sim2.055$ and $P_d/P_c\sim1.892$, both of which are close to a 2:1 MMR. To check whether the planets are truly in resonance, it is necessary to examine the critical resonant angles. These measure the locations of the planetary conjunctions relative to the pericenters of the two orbits. Two critical resonant angles exist for a 2:1 MMR between a generic planet ``1'' and planet ``2'':
\begin{align}
\phi_{12,1} &= 2\lambda_2 - \lambda_1 - \varpi_1 \\
\phi_{12,2} &= 2\lambda_2 - \lambda_1 - \varpi_2.
\end{align}
Here $\lambda$ is the mean longitude, and $\varpi$ is the longitude of periapse. For a planet pair in a 2:1 MMR, one or both of $\phi_{12,1}$ and $\phi_{12,2}$ undergo bounded oscillations about their equilibria, corresponding to a so-called ``resonant libration amplitude'' $A_{\mathrm{lib}} < 180^{\circ}$. A non-resonant planet pair has both $\phi_{12,1}$ and $\phi_{12,2}$ circulating (i.e. varying through the full $0^{\circ}-360^{\circ}$ over time). 

We performed a suite of 4,000 $N$-body integrations to check whether the planets are indeed undergoing resonant librations. The initial conditions were randomly drawn from the posterior distribution of the joint fit. For each set of initial conditions, we ran a 200 year $N$-body integration using the Wisdom-Holman \texttt{WHFast} method \citep{WisdomHolman1991, Rein2015} in texttt{rebound} \citep{Rein2012}. The orbits were assumed to be coplanar in all cases. Since the joint fit assumed circular orbits, we gave tiny initial eccentricities $e\sim10^{-7}$ (otherwise the periapses are undefined) and randomized $\varpi\sim \mathrm{Unif}(0^{\circ},360^{\circ})$. We tracked four critical resonant angles (two for each of the b/c and c/d pairs): $\phi_{bc,b}$, $\phi_{bc,c}$, $\phi_{cd,c}$, and $\phi_{cd,d}$. We also measured their libration amplitudes using a simple half peak-to-peak amplitude, $A_{\mathrm{lib}} = (\max \phi - \min \phi)/2$. We computed this only for the second half of the integration to give the system time to stabilize.

We found that some initial conditions correspond to librating solutions, and they are more common for the b/c pair. For $\phi_{bc, b}$, about $\sim$$70\%$ of initial conditions have $A_{\mathrm{lib}} < 180^{\circ}$. For $\phi_{cd, c}$, it is only $\sim$$9\%$. The other two critical resonant angles have no cases with $A_{\mathrm{lib}} < 180^{\circ}$, but it is not required that all critical resonant angles librate to be in resonance. Figure \ref{fig:critical angles} shows short segments of the time evolution of the critical resonant angles for examples where the planet pairs are resonant. Such examples are illustrative in the sense that the libration amplitude is never very small. We found that the minimum libration amplitude was $\sim23^{\circ}$ for $\phi_{bc, b}$ and $\sim43^{\circ}$ for $\phi_{cd, c}$. In summary, we find that the joint fit is consistent with some resonant solutions assuming that the orbits are very near circular. However, considering larger eccentricities would likely decrease the resonant fractions, so this could be reevaluated with more data in future work. 

\subsection{Interior Characterisation}
\label{sec:internal_structure}
\begin{figure}
\begin{center}
\includegraphics[width=\linewidth]{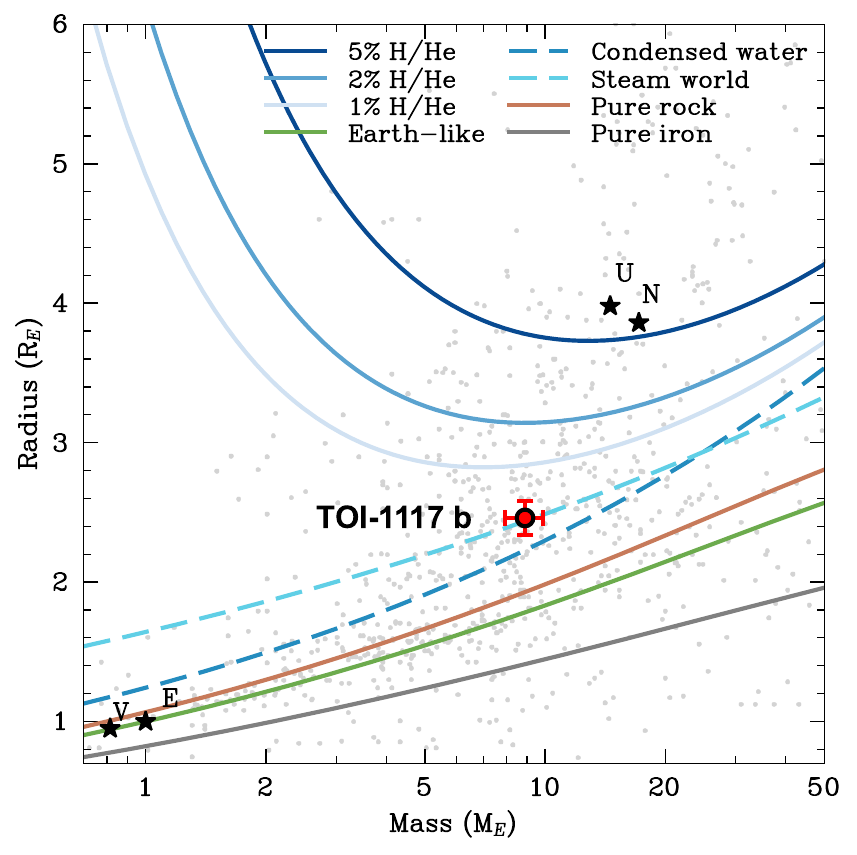}
\caption{Mass-radius relation of exoplanets from the NASA Exoplanet Archive (\url{https://exoplanetarchive.ipac.caltech.edu}). The solid blue lines show the mass-radius relation for various hydrogen envelope compositions \citep{Chen16:atmosphere-model}, the solid brown line for a pure rock composition, green line for an Earth-like composition, and gray line for a pure iron composition \citep{Zeng_2019}. The dashed cyan line shows the mass-radius relation for a steam world and the darker blue dashed line for a condensed water world compositions \citep{Venturini24} with the insolation flux of TOI-1117\,b.}
\label{fig:M_R}
\end{center}
\end{figure}

From the three planets detected in the TOI-1117 system, only TOI-1117\,b has strong constraints on both the radius and the mass, allowing us to investigate its internal structure.
%
%
As shown in Figure~\ref{fig:M_R}, TOI-1117\,b is located between the Earth-like composition line and the pure water composition line in the mass-radius (M-R) diagram.
%
%

%

%
As discussed in \citet{Otegi_2020b}, in the region in the M-R relation between the Earth-like and the pure water line, the inferred planetary composition is highly degenerate.
Furthermore, even when reducing the observational uncertainties, this degeneracy does not decrease. TOI-1117\,b falls into this region, and as a result, constraining the composition of the planet is not an easy task. 

\begin{figure}
\begin{center}
\includegraphics[width=\linewidth]{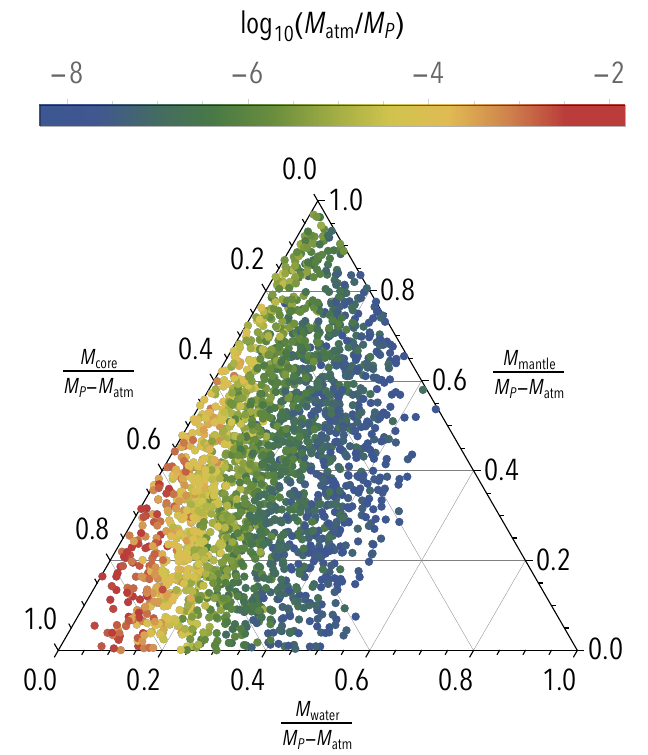}
\caption{\bf Sample of interior structures consistent with the observational constraints of TOI-1117\,b.}
\label{fig:interior_structure_posterior_sample}
\end{center}
\end{figure}

However, in order to identify the possible solutions for the internal structure of the planet, we assume various interiors  using a four-layered interior model that includes an iron core, a rocky mantle, a water layer, and a H-He atmosphere \citep{Dorn_2017, Otegi_2021}.
For the core, mantle, water layer, and atmosphere, we employ the equations of state of \citet{Hakim_2018}, \citet{Sotin_2007}, \citet{Haldemann_2020}, and \citet{Chabrier_2019}, respectively.
Using a nested sampling algorithm \citep{Buchner_2014}, we investigated the range of layer masses that would fit the observed radius and mass of TOI-1117\,b. In addition to the layer masses, we vary the planet's age, and the elemental ratios of [Fe/Si] and [Mg/Si] in the mantle. The measured age and [Mg/Si] of the host star serve as the mean values of our priors, with no assumed iron pollution of the core. However, our approach allows for deviations from stellar abundances, as star and planet compositions need not be identical \citep{Adibekyan_2021}.
Figure \ref{fig:interior_structure_posterior_sample} shows a sample of interior structures of this model that are consistent with the observational constraints of TOI-1117\,b. 
As expected, we find that a wide range of compositions and  interior structures are consistent with the observational constraints of TOI-1117\,b. These structures range from a roughly Earth-like composition with a large atmosphere (red dots) to sizable oceans without any atmosphere at all (blue dots). However, the inferred   mass fraction of water never exceeds 0.6, leaving the lower right region of Figure \ref{fig:interior_structure_posterior_sample} unpopulated. This is because higher values would lead to radii that exceed TOI-1117\,b’s observed radius. Similarly, a pure iron planet is  too dense and compact compared to TOI-1117\,b, leaving the lower left region of Figure \ref{fig:interior_structure_posterior_sample} empty.
It should be noted that the interior structure model only considers pure H-He atmospheres and neglects any pollution with heavier elements which would contract the atmosphere (e.g., \citealt{Lozovsky_2018}). Furthermore, it is still unknown under what conditions planets are differentiated into distinct compositional layers. For example, water could dissolve within the mantle of the planet \citep[e.g.,][]{Dorn_2021} or be present in the atmosphere \citep[e.g.,][]{Mol_Lous_2022}. In addition, for larger planetary masses, layers may start to intermix \citep{Helled_2017, Bodenheimer_2018, Knierim_2024}. As a result, it should be noted that the inferred bulk compositions are based on a rather simplistic modeling approach.

Additional data (e.g., atmospheric composition) or the enforcement of additional artificial constraints (e.g., assuming that the planet is devoid of water or of an atmosphere as presented in Appendix \ref{sec:interior_structure_appendix}) might help to reduce uncertainties. 
However, because of the challenges planets like TOI-1117\,b pose to interior characterization, they are an interesting case study for atmospheric measurements (i.e., identifying the existence of an atmosphere and its composition).
However, linking such measurements to the planet’s interior is challenging. First, it is unclear how atmospheric spectra is translated to heavy-element mass fractions. Second, the measured atmospheric metallicity does not necessarily reflect the planetary  bulk composition \citep{Muller2023, Knierim_2024}. Nonetheless, such constraints have the potential to further constrain the planetary interior and its primordial state \citep{Knierim_2025}.
In addition, planet formation and evolution models could be used to predict the planetary composition for such planets. 

\subsection{Photoevaporation history}
\label{sec:photoevaporation}

\begin{figure*}
    \centering
    \includegraphics[width=0.495\textwidth,trim={0 0 15cm 0},clip]{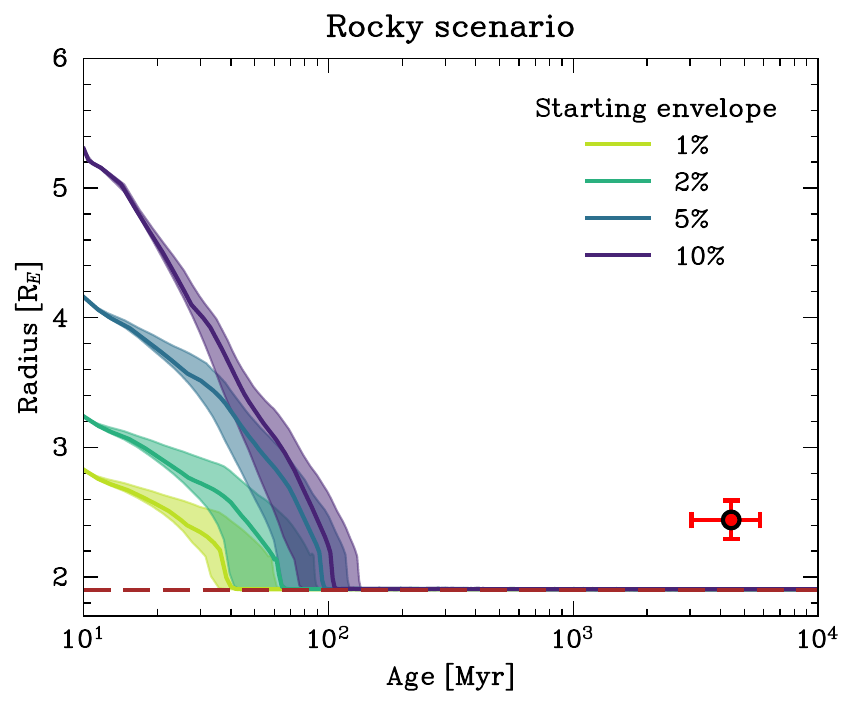}
    \includegraphics[width=0.495\textwidth,trim={0 0 15cm 0},clip]{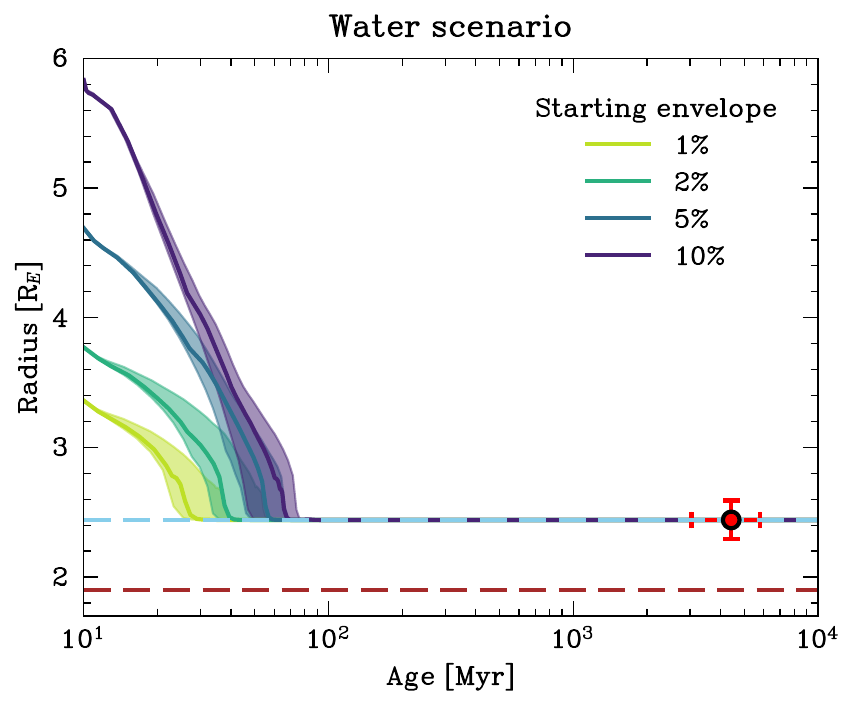}
    \caption{
        Evolution of the radius of TOI-1117\,b with a range of starting envelope mass fractions assuming rocky-H/He internal structure with no water (left panel), and a rock-water composition with no gaseous envelope (right panel).
        The planet's radius as observed in the present day is shown as a red circle, and the predicted radius of its rocky core is shown as a dashed brown line whilst the radius of the ocean layer is shown as a dashed blue line.
        The solid lines represent the evolution of the gaseous atmosphere, assuming that the planet formed in-situ and with different assumptions for the starting H/He mass fraction, ranging from 1\% to 10\%.
        The shaded regions about each solid line represent the evolution under a variety of X-ray irradiation histories expected for a 1\,M$_\odot$ star.
    }
    \label{fig:planet_evolution}
\end{figure*}

TOI-1117\,b is located very close to its star at the edge of the Neptunian desert, with a short period of 2.23 days, and so we expect it to be highly irradiated.
Its density, however, is only 3.3\,g\,cm$^{-3}$, suggesting the presence of volatiles, and thus consistent with hosting a gaseous atmosphere. This is puzzling given the strong X-ray and extreme ultra-violet (EUV) flux the planet is expected to receive, which should have stripped the planet of any gaseous atmosphere via photoevaporation \citep[e.g.][]{Owen17:radius-valley, Owen2018}.

We thus studied the past evolution of TOI-1117\,b under photoevaporation in order to determine whether a H/He-rich envelope on the planet would survive to the present day, or whether alternative compositional models for its internal structure that are free of H/He are required in order to explain its low density.

Following the internal structure analysis presented in Section\,\ref{sec:internal_structure}, we considered two scenarios for the composition of TOI-1117\,b: (1) the {\em rocky} scenario, where the planet has a rocky core surrounded by a pure H/He envelope, and (2) the {\em water} scenario, where the planet has a rocky core and a water-rich layer with no H/He envelope.

In the rocky scenario, the planet hosts a H/He envelope consisting of 0.33\% of its mass with a thickness of 0.7\,R$_\oplus$. In the water scenario, the planet hosts no H/He atmosphere and instead contains a water-rich layer consisting of 33\% of the planet's mass with a thickness of 0.98\,R$_\oplus$.

We simulated the past evaporation history of TOI-1117\,b under a variety of X-ray irradiation histories and initial H/He fractions.
We adopted the {\tt photoevolver} code \citep{Fernandez23:photoevolver} to perform the simulations, which combines models for (1) the stellar X-ray evolution, (2) the internal structure of the planet, and (3) atmospheric escape, to study the evolution of the planet's atmosphere under X-ray irradiation across its lifetime.

We adopted the stellar evolution models by \citet{Johnstone21:xray-evolution} to model the past evolution of the star's X-ray emission and thus the high energy environment of the planet throughout its life.
In our simulations, we accounted for the variety of past X-ray emission histories predicted by \citet{Johnstone21:xray-evolution}, between the 5th and 95th percentiles in their models, which reflects the diversity of spin periods and X-ray activities observed on main sequence stars at a given age.

We additionally adopted the envelope structure model by \citet{Chen16:atmosphere-model}, based on the {\it MESA} code \citep{Paxton11:mesa}, to model the thermal evolution of the gaseous envelope, as well as the mass loss model by \citet{Kubyshkina18:mass-loss-model}, based on hydrodynamic simulations of escaping atmospheres on a grid of planet models with 1--39\,M$_\oplus$ and 1--10\,R$_\oplus$.

We evolved both the rocky and water scenarios with a number of starting H/He mass fractions ranging from 1\% to 10\%. We ran the simulations from the age of 10 Myr to 10 Gyr with a variable time-step no longer than 5 Myr determined with the 4th order Runge-Kutta algorithm \citep{shampine1986runge}.

Our results are shown on Figure\,\ref{fig:planet_evolution}, where we plotted the evolution of TOI-1117\,b under photoevaporation for the rocky scenario (left panel) and the water scenario (right panel). We found that, in all cases, the planet is fully stripped of its primordial H/He envelope within the first 100\,Myr of age.

Our simulations therefore suggest the planet should not host a H/He envelope in the present day, which, providing that TOI-1117\,b formed in-situ, rules out the rocky scenario and instead favours the water scenario for TOI-1117\,b.

Moreover, in our simulations of the water scenario we did not consider mixing between the H/He and the water-rich layers. A higher degree of mixing would increase the metallicity of the gaseous atmosphere, enhancing its molecular weight and thus making it more resistant to photoevaporation \citep{OwenJackson12:xray-evap}. As a result, it is possible that, under this scenario, a thin H/He atmosphere with a high metallicity may have survived for longer.

\section{Discussion}
\label{sec:Discussion}
\begin{figure*}
    \includegraphics[width=\textwidth,trim={0 5.5cm 0 5.5cm},clip]{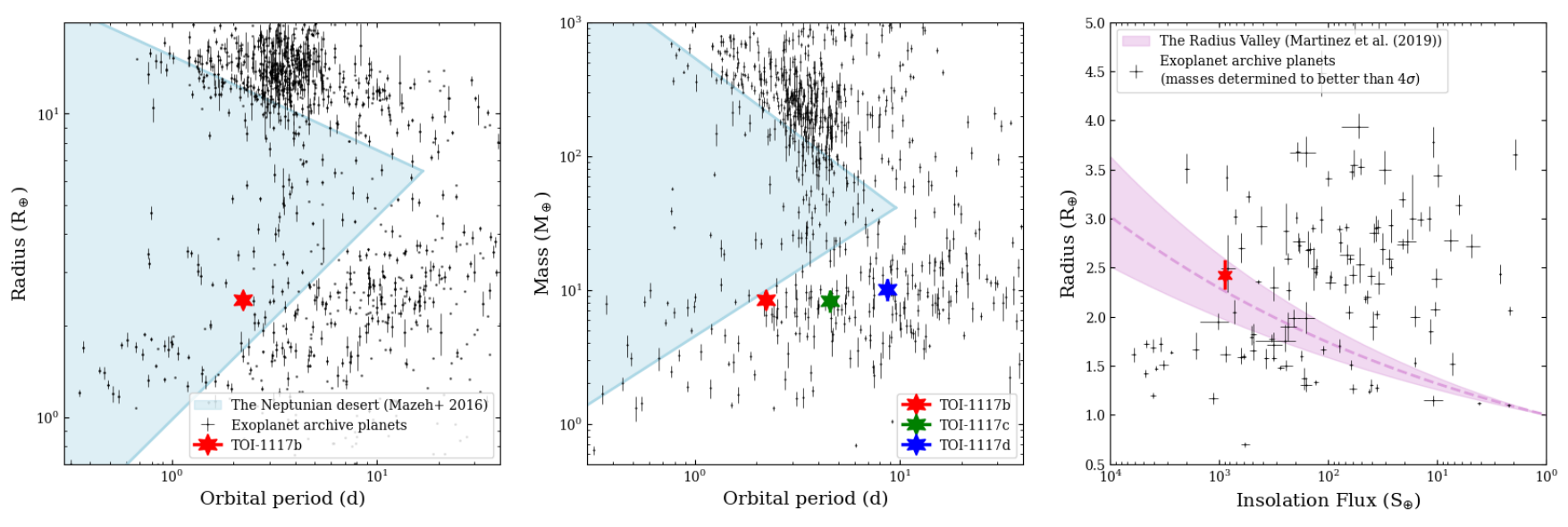}
    \caption{Left: Plot of all NASA Exoplanet Archive planets with masses determined to better than 4$\sigma$ in radius-period space showing TOI-1117\,b (red star) to fall within the Neptunian Desert (pale blue region). Middle: Mass-period plot showing TOI-1117\,b (red), c (green), and d (blue) and the Neptunian Desert. Right: Plot of radius  against insolation flux to show TOI-1117\,b to be within the Radius Valley (pale pink region).}
 \label{fig:PRPM}
\end{figure*}

This section uses the results of analyses presented in Section~\ref{sec:Analysis} to discuss potential formation and evolution mechanisms for the TOI-1117 system. TOI-1117\,b is a rare example of a short-period sub-Neptune, residing in both the Radius Valley and on the edge of the Neptunian Desert (Figure~\ref{fig:PRPM}). Current theories struggle to explain the formation of such planets.

There is a case for in situ formation at the disc inner edge where accreted gas and dust has higher densities and metallicities than further out in the disc \citep{Ogihara}, enhancing the molecular weight of the atmosphere and thus making it more resistant to photoevaporation. However, it is unlikely that TOI-1117\,b formed in situ at 0.033AU because photoevaporation analysis (Section~\ref{sec:photoevaporation}) shows TOI-1117\,b is likely to host water and water accretion is expected to occur beyond the snow line \citep{Dawson2018}.

Another theory is that short-period Neptunes follow the same early stages as hot Jupiters -- forming in outer regions and migrating inwards -- but then undergo photoevaporation or core-powered mass loss \citep{Gupta2020}, removing significant mass from the gas envelope \citep{Dawson2018}. A water-rich or metal-rich layer may be less susceptible to photoevaporation and could therefore explain how a thin gas envelope was retained. Popular theories for the migration mechanism include disk migration \citep{Goldreich1979,Lin1996,Baruteau2014} and high-eccentricity migration \citep{WuMurray2003,FordRasio2008,Dawson2018}, both of which can explain the formation of close-in stable and circularised systems like TOI-1117.

The link between hot Jupiters and hot Neptunes is based on currently known exoplanets in the Neptunian Desert, which are mostly single-planet systems found around high-metallicity stars \citep{Dong2017}. However, TOI-1117\,b is more complex as it has 2 outer sub-Neptunes, also in stable, circularised orbits. The three planets appear to have similar masses and evenly spaced separations, lending themselves to a three 'peas-in-a-pod' scenario. This term refers to a pattern, first recognized by \citet{Weiss2018}, in which planets in multi-planet system tend to have similar sizes, masses, and orbital spacings. \citet{Gilbert2020} quantified the notion of peas-in-a-pod with two metrics; Mass Partitioning, $Q\in[0,1]$, which describes variations in planetary masses, and Gap Complexity, $C\in[0,1]$, which describes regularities in orbital separations. The TOI-1117 system has $Q=0.004$ and $C=0.005$, both consistent with 0, indicating a highly uniform system in both mass and spacing. The Radius Valley is more pronounced for multi-planet systems with low gap complexity, $C<0.165$, suggesting that multi-planet systems end up in the Radius Valley because of strong dynamical interactions resulting from irregular spacings \citep{Rice2024}. A more quiescent concept is needed to explain the system architecture of TOI-1117 as the planets are regularly space, likely co-orbital, and near-resonant (Section~\ref{sec:Analysis}).

One such alternative is that the origin of TOI-1117\,b is more closely-linked to the formation of ultra-short-period ($P<1$~day) rocky Earths, which commonly reside in multi-planet systems \citep{Sanchis-Ojeda2014}. They are thought to form via low-eccentricity migration. \citet{Pu2019} performed population synthesis of 3-planet systems formed ex situ with low-eccentricity migration. They found that protoplanets experience less orbital decay, generating systems with short-period, $P>1$ day, mini-Neptune inner planets and slightly more massive outer planets. Furthermore, the study found that synthesised systems generating larger, short-period inner planets also had significantly smaller mutual inclinations and were 46\% more likely to have transiting companion planets. 

Future studies measuring the orbital decay rate of planets in the Neptunian Desert may be useful in further understanding their formation and evolution.

\section{Conclusions}
\label{sec:Conclusions}

This paper collated photometric and spectroscopic observations of TOI-1117 in the search for planets in the Neptunian Desert. TOI-1117 is a high-metallicity, solar-type star, with an age estimate of $4.42\pm1.50$ Gyr. TOI-1117 hosts three sub-Neptune-sized planets on short-period orbits. Only non-eccentric system architectures were found to be stable beyond 4 Gyr, showing that fitting to observations can produce results which are not dynamically viable and stability analyses can be used to better constrain orbital parameters than the joint fit alone.

Resonance analysis of TOI-1117 presents evidence of a near 2:1 resonance between TOI-1117\,b and c, with 70\% of solutions librating. Only 9\% of solution for the c/d pair were librating so it is unlikely that this is part of a resonant chain.  
    
The inner planet, TOI-1117\,b, is on the edge of the Neptunian Desert and in the Radius Valley. The planetary parameters of TOI-1117\,b are consistent with a range of internal structure models, from an Earth-like composition to an ocean world with no atmosphere. Given that TOI-1117\,b has a low density of 3.30~g~cm$^{-3}$, we expect it to host a gaseous atmosphere; however, given the proximity to TOI-1117, we expect any atmosphere to have been stripped via photoevaporation. This contradiction could be explained by the presence of a water-rich layer. Atmospheric characterisation should be performed, ideally with JWST, to provide information on the evolution of TOI-1117\,b. The ESM for TOI-1117\,b is 11.7 which is high enough for JWST (ESM>10) but not as high as other short-period Neptunes (mean ESM = 21.6) \citep{Kempton2018}. Confirming the presence of a steam atmosphere or ocean and its metallicity would be vital information to understand the current state of the atmosphere \citep{OwenJackson12:xray-evap}.

TOI-1117’s multi-planet system does not conform with current formation theories. For TOI-1117\,b to be water-rich, it likely formed beyond the snow line and migrated inwards. The circular eccentricities support disk migration over high-eccentricity migration theories \citep{Dawson2018}. Moreover, the multi-planet nature of the system favours low-eccentricity migration \citep{Pu2019}. Giant planets on longer orbits could be present and need to be investigated with future radial velocity measurements.

\section{Acknowledgements}

Funding for the TESS mission is provided by NASA's Science Mission Directorate. We acknowledge the use of public TESS data from pipelines at the TESS Science Office and at the TESS Science Processing Operations Center (SPOC). Resources supporting this work were provided by the NASA High-End Computing (HEC) Program through the NASA Advanced Supercomputing (NAS) Division at Ames Research Center for the production of the SPOC data products. KAC and CNW acknowledge support from the TESS mission via subaward s3449 from MIT. This research has made use of the Exoplanet Follow-up Observation Program website, which is operated by the California Institute of Technology, under contract with the National Aeronautics and Space Administration under the Exoplanet Exploration Program. This paper includes data collected by the TESS mission that are publicly available from the Mikulski Archive for Space Telescopes (MAST).
This research has made use of the Exoplanet Follow-up Observation Program (ExoFOP; DOI: 10.26134/ExoFOP5) website, which is operated by the California Institute of Technology, under contract with the National Aeronautics and Space Administration under the Exoplanet Exploration Program.
This work makes use of observations from the LCOGT network. Part of the LCOGT telescope time was granted by NOIRLab through the Mid-Scale Innovations Program (MSIP). MSIP is funded by NSF.
This paper makes use of data from the MEarth Project, which is a collaboration between Harvard University and the Smithsonian Astrophysical Observatory. The MEarth Project acknowledges funding from the David and Lucile Packard Fellowship for Science and Engineering, the National Science Foundation under grants AST-0807690, AST-1109468, AST-1616624 and AST-1004488 (Alan T. Waterman Award), the National Aeronautics and Space Administration under Grant No. 80NSSC18K0476 issued through the XRP Program, and the John Templeton Foundation.
Some of the observations in this paper made use of the High-Resolution Imaging instrument Zorro and were obtained under Gemini LLP Proposal Number: GN/S-2021A-LP-105. Zorro was funded by the NASA Exoplanet Exploration Program and built at the NASA Ames Research Center by Steve B. Howell, Nic Scott, Elliott P. Horch, and Emmett Quigley. Zorro was mounted on the Gemini South telescope of the international Gemini Observatory, a program of NSF’s OIR Lab, which is managed by the Association of Universities for Research in Astronomy (AURA) under a cooperative agreement with the National Science Foundation. on behalf of the Gemini partnership: the National Science Foundation (United States), National Research Council (Canada), Agencia Nacional de Investigación y Desarrollo (Chile), Ministerio de Ciencia, Tecnología e Innovación (Argentina), Ministério da Ciência, Tecnologia, Inovações e Comunicações (Brazil), and Korea Astronomy and Space Science Institute (Republic of Korea).
This research was funded in part by the UKRI, (Grants ST/X001121/1, EP/X027562/1). Based on observations collected at the European Southern Observatory under ESO programme 108.21YY.001 (PI Armstrong). 
A.C.-G. and J.L.-B. are funded by the Spanish Ministry of Science and Universities (MICIU/AEI/10.13039/501100011033/) and NextGenerationEU/PRTR grants PID2019-107061GB-C61 and CNS2023-144309.  
NCS is funded by the European Union (ERC, FIERCE, 101052347). Views and opinions expressed are however those of the author(s) only and do not necessarily reflect those of the European Union or the European Research Council. Neither the European Union nor the granting authority can be held responsible for them. This work was supported by FCT - Fundação para a Ciência e a Tecnologia through national funds and by FEDER through COMPETE2020 - Programa Operacional Competitividade e Internacionalização by these grants: UIDB/04434/2020; UIDP/04434/2020.
S.G.S acknowledges the support from FCT through Investigador FCT contract nr. CEECIND/00826/2018 and POPH/FSE (EC).
Part of this work has been carried out within the framework of the National Centre of Competence in Research PlanetS supported by the Swiss National Science Foundation under grants \texttt{51NF40\_182901} and \texttt{51NF40\_205606}.
This work made use of \texttt{tpfplotter} by J. Lillo-Box (publicly available in \url{www.github.com/jlillo/tpfplotter}), which also made use of the python packages \texttt{astropy}, \texttt{lightkurve}, \texttt{matplotlib} and \texttt{numpy}.This work made use of \texttt{TESS-cont} (\url{https://github.com/castro-gzlz/TESS-cont}), which also made use of \texttt{tpfplotter} \citep{Aller2020} and \texttt{TESS-PRF} \citep{Bell2022}.
The authors would like to acknowledge the University of Warwick Research Technology Platform SCRTP for assistance in the research described in this paper.

\section{Data Availability}

The TESS data are available from MAST, at \url{https://mast.stsci.edu/portal/Mashup/Clients/Mast/Portal.html}. The other photometry from LCOGT and M-Earth-South, and the speckle imaging from SOAR and Gemini are available for public download from the ExoFOP-TESS archive at \url{https://exofop.ipac.caltech.edu/tess/target.php?id=295541511}. The full HARPS RV data products are publicly available from the ESO archive, at \url{https://archive.eso.org/wdb/wdb/eso/eso_archive_main/query}. The \texttt{rebound} simulation code and the \texttt{exoplanet} modeling code, parameter analysis, and plotting can be made available upon reasonable request to the author.

\bibliographystyle{mnras}
\bibliography{main}

\appendix

\section{HARPS RVs}

\begin{table}
 \vspace{-1em}
 \caption{HARPS radial velocities of TOI-1117. Rows in bold indicate data removed before joint fits were performed.}
 \vspace{-1em}
 \label{tab:RVs}
  \begin{tabular}[t]{ccccccc}
  \hline
   Time & RV & RV Error & FWHM & BIS & S-Index\\ 
   RJD & ms$^{-1}$ & ms$^{-1}$ & ms$^{-1}$ & ms$^{-1}$ &\\
   \hline
   59736.85978 & -24887.7 & 2.8 & 7199.2 & -56.4 & 0.16\\
   59737.78413 & -24893.3 & 2.7 & 7192.6 & -56.6 & 0.17\\
   59759.71327 & -24894.5 & 3.2 & 7204.1 & -59.5 & 0.16\\
   59760.68907 & -24890.4 & 2.1 & 7198.1 & -61.1 & 0.16\\ 
59761.65155 & -24898.4 & 2.6 & 7206.0 & -54.9 & 0.16\\
59767.73875 & -24888.3 & 3.4 & 7207.2 & -71.2 & 0.16\\ 
59769.64222 & -24890.1 & 4.8 & 7206.9 & -62.9 & 0.15\\
59810.60030 & -24898.6 & 2.8 & 7211.2 & -51.5 & 0.16\\ 
59811.64316 & -24892.2 & 2.6 & 7217.0 & -57.4 & 0.15\\ 
59812.61762 & -24893.6 & 3.2 & 7196.8 & -53.7 & 0.15\\ 
59813.57509 & -24886.2 & 2.3 & 7213.5 & -60.4 & 0.15\\
59814.57321 & -24887.3 & 4.2 & 7194.9 & -74.5 & 0.18\\ 
59821.58534 & -24890.9 & 2.6 & 7196.8 & -59.3 & 0.15\\ 
59822.55920 & -24884.0 & 3.2 & 7220.1 & -66.4 & 0.15\\
59823.59574 & -24874.2 & 6.7 & 7196.0 & -57.6 & 0.14\\ 
59824.55816 & -24892.6 & 4.8 & 7199.8 & -66.1 & 0.11\\ 
59825.53833 & -24891.7 & 5.8 & 7191.8 & -61.5 & 0.13\\ 
59834.59773 & -24894.4 & 2.9 & 7193.2 & -69.4 & 0.15\\ 
59835.61281 & -24896.8 & 4.2 & 7193.9 & -66.7 & 0.19\\
59836.59946 & -24888.7 & 6.9 & 7187.9 & -72.9 & 0.20\\ 
59837.58237 & -24898.2 & 5.4 & 7207.1 & -63.1 & 0.16\\ 
\textbf{59847.53889} & \textbf{-24907.8} & \textbf{17.0} & \textbf{7148.4} & \textbf{-79.0} & \textbf{0.26}\\ 
59848.59674 & -24889.3 & 3.5 & 7201.7 & -54.7 & 0.16\\
59849.57487 & -24883.2 & 10.5 & 7173.8 & -85.7 & 0.17\\ 
59850.54555 & -24886.7 & 3.2 & 7198.8 & -55.7 & 0.17\\ 
59852.56573 & -24898.6 & 2.1 & 7194.5 & -60.5 & 0.16\\ 
59853.51610 & -24898.9 & 2.6 & 7210.6 & -63.0 & 0.16\\ 
59854.56000 & -24895.8 & 3.9 & 7211.4 & -76.0 & 0.15\\ 
59855.54627 & -24895.1 & 3.6 & 7190.0 & -59.5 & 0.15\\ 
59856.55137 & -24895.5 & 4.6 & 7194.4 & -51.4 & 0.15\\ 
59857.52516 & -24896.5 & 5.6 & 7207.3 & -56.7 & 0.18\\
59857.58466 & -24900.3 & 5.7 & 7204.2 & -41.2 & 0.09\\ 
59859.51178 & -24896.0 & 3.2 & 7214.5 & -52.3 & 0.17\\ 
59860.51947 & -24894.3 & 2.7 & 7205.0 & -56.7 & 0.16\\
59861.51362 & -24893.8 & 4.7 & 7218.5 & -53.4 & 0.13\\
59862.49640 & -24898.5 & 6.2 & 7227.6 & -66.2 & 0.17\\ 
59863.55270 & -24885.3 & 3.0 & 7223.3 & -65.0 & 0.15\\ 
59864.51511 & -24894.1 & 2.5 & 7208.3 & -57.7 & 0.16\\ 
60062.80049 & -24903.7 & 2.3 & 7213.4 & -58.6 & 0.17\\ 
60062.88717 & -24903.1 & 2.2 & 7211.9 & -57.4 & 0.15\\ 
60063.82159 & -24891.8 & 1.8 & 7212.2 & -62.9 & 0.17\\ 
60064.91698 & -24896.1 & 4.0 & 7212.5 & -72.1 & 0.19\\ 
60065.82540 & -24885.6 & 2.3 & 7202.2 & -67.6 & 0.14\\ 
60066.69479 & -24899.3 & 2.6 & 7213.2 & -64.4 & 0.14\\ 
60067.73258 & -24901.5 & 2.9 & 7210.7 & -57.2 & 0.17\\ 
60068.74214 & -24896.5 & 5.0 & 7216.0 & -61.0 & 0.14\\ 
60069.78603 & -24894.1 & 2.5 & 7228.2 & -67.2 & 0.16\\ 
60069.87602 & -24893.9 & 2.4 & 7205.6 & -66.1 & 0.17\\ 
60070.77632 & -24885.5 & 2.6 & 7208.0 & -51.5 & 0.16\\ 
60070.85217 & -24890.9 & 2.3 & 7207.8 & -57.3 & 0.16\\ 
\textbf{60071.71331} & \textbf{-24926.5} & \textbf{19.0} & \textbf{7406.1} & \textbf{-1.4} & \textbf{0.15}\\
60071.80771 & -24901.8 & 2.7 & 7224.2 & -54.6 & 0.15\\
60072.84435 & -24892.5 & 2.3 & 7200.0 & -75.3 & 0.17\\ 
60073.91702 & -24888.7 & 2.6 & 7214.7 & -52.4 & 0.17\\ 
60074.73993 & -24881.5 & 2.0 & 7195.7 & -60.9 & 0.16\\ 
60074.90546 & -24878.8 & 2.4 & 7226.3 & -57.8 & 0.18\\ 
60078.63920 & -24896.5 & 3.0 & 7205.5 & -58.4 & 0.16\\ 
60079.83830 & -24889.7 & 2.6 & 7214.9 & -64.8 & 0.16\\ 
60080.78349 & -24896.0 & 2.7 & 7210.6 & -66.0 & 0.16\\ 
60082.81198 & -24898.9 & 2.7 & 7214.8 & -60.0 & 0.15\\ 
60083.82408 & -24889.0 & 3.6 & 7220.8 & -65.8 & 0.15\\
60084.87544 & -24900.6 & 2.9 & 7208.1 & -65.6 & 0.16\\
60085.85277 & -24896.4 & 5.6 & 7223.9 & -57.3 & 0.15\\
60086.82361 & -24912.3 & 4.1 & 7180.2 & -75.0 & 0.17\\ 
\end{tabular}
\end{table}
\begin{table}
 \vspace{-1em}
\begin{tabular}[t]{cccccc}
60087.73551 & -24900.3 & 2.6 & 7217.6 & -67.2 & 0.15\\ 
60088.84240 & -24896.0 & 2.5 & 7208.7 & -66.0 & 0.17\\ 
60089.71963 & -24896.4 & 2.8 & 7202.6 & -66.6 & 0.15\\ 
60090.73730 & -24892.3 & 3.6 & 7204.9 & -54.2 & 0.15\\ 
60106.84845 & -24898.3 & 3.5 & 7241.8 & -61.3 & 0.18\\ 
60107.68405 & -24898.6 & 5.4 & 7221.0 & -59.0 & 0.13\\ 
60108.74752 & -24904.7 & 4.3 & 7219.6 & -68.6 & 0.14\\ 
60109.73667 & -24900.8 & 4.4 & 7217.7 & -59.6 & 0.17\\ 
60110.72255 & -24887.1 & 2.5 & 7212.5 & -59.1 & 0.16\\ 
60110.77465 & -24886.8 & 3.5 & 7193.9 & -58.6 & 0.16\\ 
60111.77166 & -24903.7 & 3.2 & 7211.7 & -61.7 & 0.18\\ 
60112.72406 & -24890.9 & 2.6 & 7211.6 & -64.0 & 0.16\\
60113.78188 & -24904.4 & 2.6 & 7203.5 & -69.9 & 0.17\\ 
60124.65192 & -24888.1 & 2.9 & 7223.8 & -51.8 & 0.16\\ 
60124.80450 & -24887.6 & 2.9 & 7209.1 & -64.4 & 0.16\\ 
60126.61935 & -24889.3 & 2.4 & 7219.9 & -62.0 & 0.16\\ 
60126.82864 & -24897.6 & 2.5 & 7221.7 & -58.2 & 0.16\\ 
60127.62056 & -24897.0 & 1.8 & 7211.5 & -64.7 & 0.17\\ 
60127.85907 & -24896.2 & 3.2 & 7219.0 & -70.1 & 0.17\\ 
60128.56051 & -24893.1 & 4.1 & 7206.6 & -56.8 & 0.13\\
60128.80848 & -24897.7 & 4.1 & 7203.5 & -61.4 & 0.16\\ 
60129.56661 & -24901.0 & 3.6 & 7212.1 & -60.3 & 0.15\\ 
60129.66336 & -24895.2 & 3.1 & 7218.2 & -64.5 & 0.16\\ 
60130.55773 & -24889.7 & 2.8 & 7216.4 & -66.4 & 0.16\\ 
60130.72237 & -24891.4 & 3.6 & 7209.8 & -54.2 & 0.16\\ 
60131.58922 & -24901.7 & 6.1 & 7173.1 & -40.2 & 0.19\\ 
60131.80977 & -24902.5 & 4.3 & 7203.5 & -63.6 & 0.15\\ 
60132.50523 & -24892.8 & 6.9 & 7174.3 & -48.1 & 0.14\\ 
60134.76573 & -24894.1 & 7.1 & 7231.5 & -96.8 & 0.12\\ 
60139.61417 & -24893.0 & 2.9 & 7193.2 & -70.3 & 0.15\\ 
60139.76669 & -24898.8 & 2.9 & 7204.3 & -73.1 & 0.17\\ 
60140.56970 & -24901.8 & 1.8 & 7209.8 & -66.0 & 0.17\\ 
60140.67757 & -24900.7 & 2.2 & 7217.4 & -60.5 & 0.18\\ 
60141.55728 & -24892.3 & 2.9 & 7211.1 & -64.4 & 0.17\\ 
60141.74364 & -24889.5 & 2.7 & 7198.7 & -72.9 & 0.17\\ 
60142.59975 & -24893.6 & 2.6 & 7213.3 & -60.2 & 0.17\\ 
60142.71187 & -24892.2 & 3.2 & 7207.8 & -60.8 & 0.17\\ 
60171.55456 & -24898.3 & 2.9 & 7216.7 & -66.4 & 0.16\\ 
60171.64163 & -24891.6 & 2.3 & 7216.0 & -60.5 & 0.15\\ 
60172.52595 & -24889.6 & 4.7 & 7165.1 & -66.0 & 0.04\\ 
60172.58930 & -24897.2 & 3.7 & 7195.4 & -72.5 & 0.19\\ 
60173.55062 & -24908.3 & 5.0 & 7179.6 & -53.7 & 0.19\\ 
60173.62382 & -24906.1 & 4.6 & 7211.3 & -61.6 & 0.17\\ 
60174.49894 & -24893.8 & 4.9 & 7231.2 & -69.4 & 0.16\\ 
60174.56856 & -24908.2 & 4.4 & 7226.7 & -68.3 & 0.16\\ 
60175.49996 & -24895.3 & 4.0 & 7204.6 & -53.1 & 0.18\\ 
60175.54531 & -24893.2 & 5.1 & 7217.8 & -71.7 & 0.14\\ 
60178.50947 & -24897.9 & 3.9 & 7204.4 & -48.8 & 0.17\\ 
60179.50912 & -24885.8 & 3.1 & 7211.4 & -59.0 & 0.15\\ 
60192.54071 & -24898.7 & 6.2 & 7182.5 & -63.7 & 0.21\\ 
60193.51608 & -24893.2 & 4.1 & 7206.1 & -55.3 & 0.17\\ 
60196.53320 & -24903.9 & 2.8 & 7225.3 & -63.1 & 0.17\\ 
60196.60669 & -24905.3 & 4.1 & 7237.6 & -77.0 & 0.16\\ 
60208.50901 & -24891.9 & 4.5 & 7221.5 & -48.9 & 0.17\\ 
60208.61493 & -24895.8 & 3.9 & 7231.3 & -57.7 & 0.18\\ 
60209.51075 & -24900.1 & 3.9 & 7220.6 & -57.7 & 0.19\\ 
60209.62107 & -24902.3 & 3.5 & 7232.1 & -53.3 & 0.16\\ 
60210.52152 & -24893.3 & 3.3 & 7225.2 & -46.5 & 0.17\\ 
60210.61631 & -24894.4 & 3.9 & 7225.2 & -58.3 & 0.16\\ 
60211.51255 & -24887.2 & 2.6 & 7202.7 & -64.8 & 0.16\\ 
60211.61788 & -24895.6 & 3.7 & 7216.2 & -60.5 & 0.16\\ 
60212.51300 & -24890.1 & 2.1 & 7214.8 & -61.2 & 0.18\\ 
60212.61221 & -24892.0 & 3.2 & 7217.1 & -65.4 & 0.17\\ 
60213.51274 & -24892.5 & 4.9 & 7224.2 & -59.4 & 0.17\\ 
60213.62166 & -24898.6 & 8.7 & 7208.9 & -74.5 & 0.18\\ 
60214.51461 & -24890.5 & 6.1 & 7210.3 & -55.9 & 0.21\\ 
60215.50651 & -24895.3 & 7.3 & 7230.3 & -57.0 & 0.21\\ 
60215.59868 & -24903.7 & 5.3 & 7205.6 & -61.2 & 0.16\\ 
60216.53211 & -24895.5 & 2.5 & 7233.5 & -67.1 & 0.18\\ 
60216.59531 & -24919.0 & 5.5 & 7198.1 & -61.6 & 0.17\\ 
60218.55846 & -24915.9 & 3.9 & 7241.9 & -57.5 & 0.16\\
   \hline
 \end{tabular}
\end{table}

\section{Chemical Abundances}
\begin{table}
 \caption{Stellar abundances for the TOI-1117 system.}
 \label{table A1}
  \begin{tabular}{lr}
   \hline
   Parameter & Value \\
   \hline
   \\\multicolumn{2}{l}{\textbf{Chemical abundances}} \\
   {[}C/H{]} & $-0.035\pm0.026$\\
   {[}O/H{]} & $0.066\pm0.112$\\
   {[}Na/H{]} & $0.16\pm0.06$\\
   {[}Mg/H{]} & $0.014\pm0.03$\\
   {[}Si/H{]} & $0.12\pm0.03$\\
   {[}Ti/H{]} & $0.15\pm0.03$\\
   {[}Ni/H{]} & $0.13\pm0.02$\\
   {[}Cu/H{]} & $0.160\pm0.028$\\
   {[}Zn/H{]} & $0.110\pm0.043$\\
   {[}Sr/H{]} & $0.173\pm0.077$\\
   {[}Y/H{]} & $0.133\pm0.058$\\
   {[}Zr/H{]} & $0.114\pm0.018$\\
   {[}Ba/H{]} & $0.088\pm0.035$\\
   {[}Ce/H{]} & $0.120\pm0.025$\\
   {[}Nd/H{]} & $0.139\pm0.024$\\
   \hline
  \end{tabular}
\end{table}

\section{Interior Structure Model} 
\label{sec:interior_structure_appendix}
In addition to the model presented in Section~\ref{sec:internal_structure}, we can put artificial constraints on TOI-1117\,b to investigate the effect on interior structure. In detail, we investigated two additional models: first, a model where we set the water mass fraction to zero to estimate the maximum atmospheric mass fraction (model no-water), and second, a model where we set the atmospheric mass fraction to zero to estimate the maximum water mass fraction (model no-atmosphere). 
Table~\ref{tab:nested_sampling_results} shows the result of the Bayesian inference of these two models together with the model from Section~\ref{sec:internal_structure} called "free".
\begin{table}
    \centering
    \caption{Nested sampling results for TOI-1117\,b. The asterisk indicates that we fixed the value for the respective model.}
    \begin{tabular}{l l l l l l l}
        \toprule
        Model & $M_\mathrm{core}/M_p$ & $M_\mathrm{mantle}/M_p$ & $M_\mathrm{water}/M_p$ & $\log(M_\mathrm{atm}/M_p)$\\
        \midrule
        no water & $0.39 \pm 0.24$ & $0.60 \pm 0.25$ & $0^*$ & $-3.41\pm 0.81$\\
        no atmosphere & $0.24 \pm 0.15$ & $0.41 \pm 0.21$ & $0.35 \pm 0.09$ & $-\infty^*$\\
        free & $0.40 \pm 0.21$ & $0.40 \pm 0.23$ & $0.20 \pm 0.12$ & $-5.88\pm 1.47$\\
        \bottomrule
    \end{tabular}
    \label{tab:nested_sampling_results}
\end{table}

\section{Additional Figures}

\begin{figure*}
    \centering
    \includegraphics[width=\columnwidth]{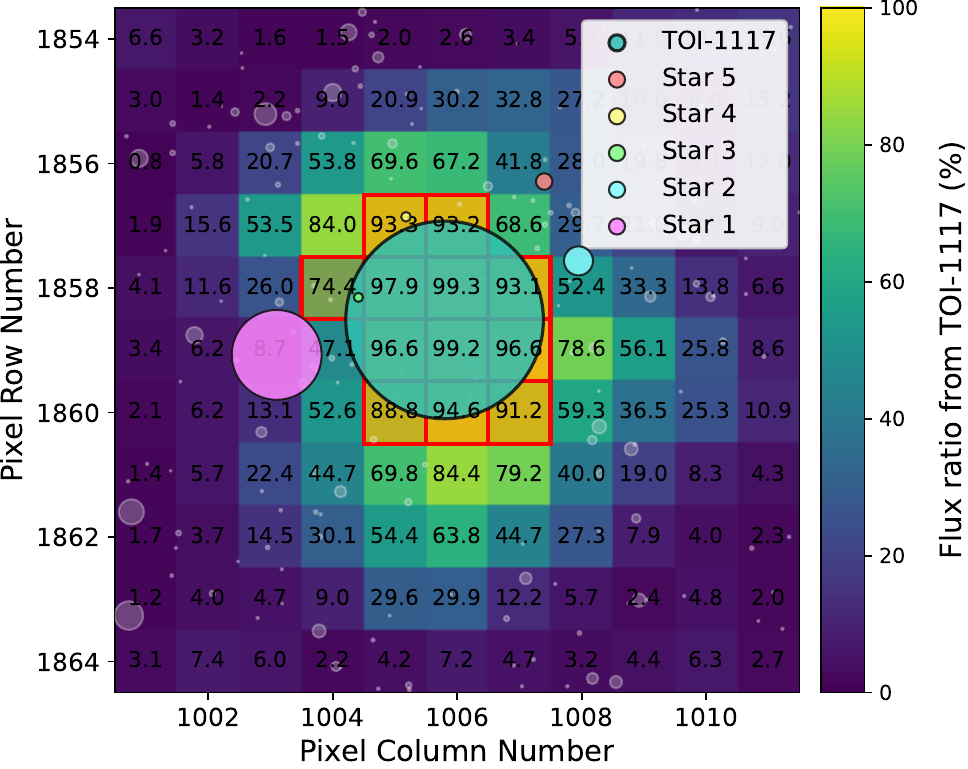}
    \includegraphics[width=\columnwidth]{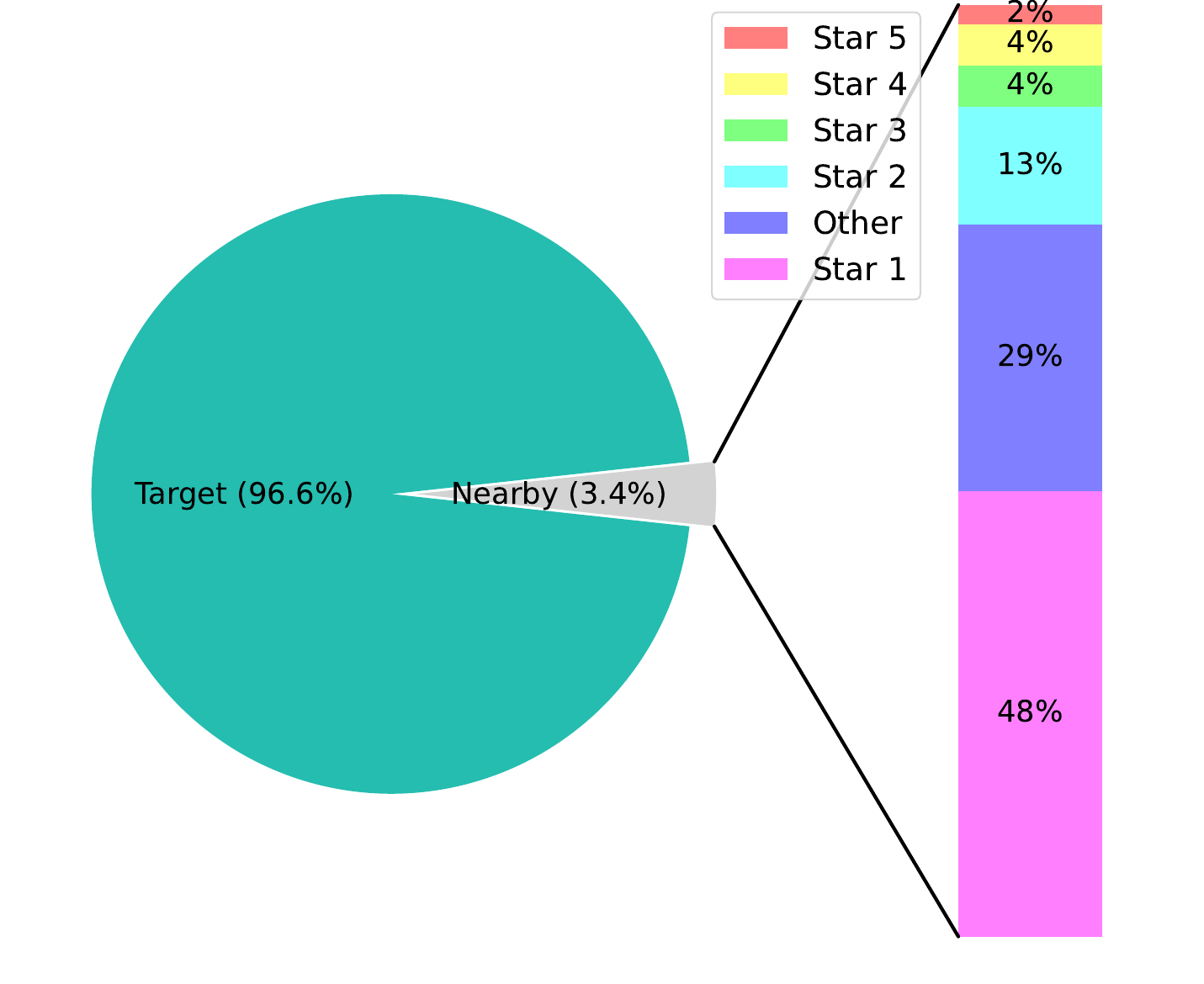}
    \caption{Nearby sources contaminating the TOI-1117 photometry. Left: TPF-shaped heatmap with the pixel-by-pixel flux fraction from TOI-1117 in S13. The red grid is the SPOC aperture. The pixel scale is 21 arcsec $\rm pixel^{-1}$. The white disks represent all the \textit{Gaia} sources, and the five sources that most contribute to the aperture flux are highlighted in different colours. The disk areas scale with the emitted fluxes. Right: Flux contributions to the SPOC aperture from the target and most contaminant stars.}
    \label{fig:tess-cont}
\end{figure*}

\begin{figure}
 \label{fig:Orbits}
    \includegraphics[width=\columnwidth]{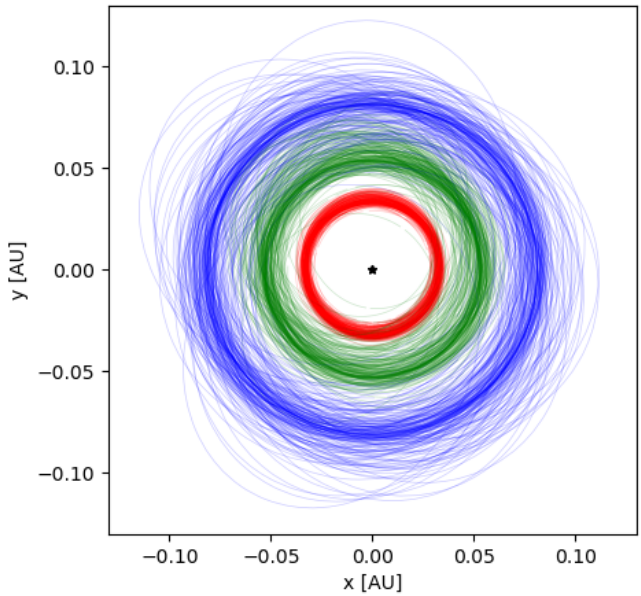}
 \caption{200 random example system architectures for TOI-1117 used in the eccentricity analysis described in Section~\ref{sec:stabilityecc}}
\end{figure}

\begin{figure}
    \includegraphics[width=\columnwidth]{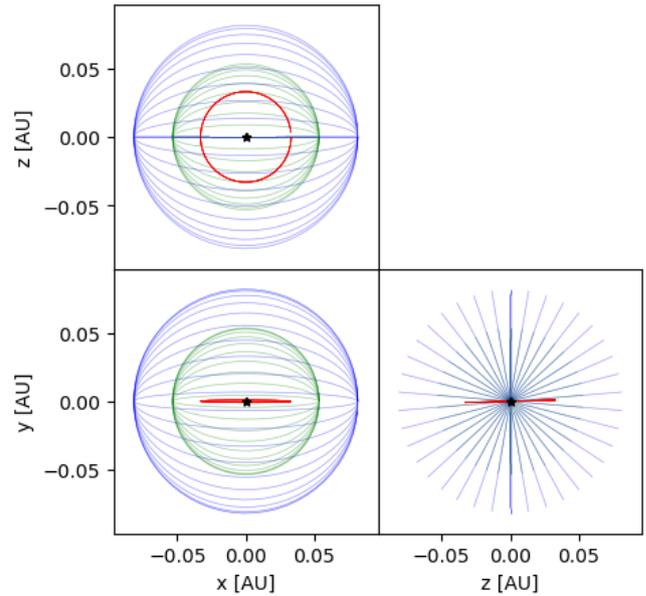}
 \caption{20 example system architectures for TOI-1117 used in the inclination analysis described in Section~\ref{sec:stabilityinc}, where the x-y plane is parallel to the plane of the sky.}
 \label{fig:inclinedOrbits}
\end{figure}

\bsp    
\label{lastpage}
\end{document}